\documentclass[twocolumn,aps,superscriptaddress,multicol,amsmath,amssymb]{revtex4-2}

\makeatletter
\newcommand*{\rom}[1]{\expandafter\@slowromancap\romannumeral #1@}
\makeatother
\usepackage{xcolor}
\usepackage{color}
\usepackage{graphicx}
\usepackage{hyperref}
\usepackage{amsmath,amssymb}
\usepackage{epstopdf}
\usepackage[font=small,labelfont=bf,justification=justified]{caption}
\usepackage[font=small,labelfont=bf,justification=justified]{subcaption}

\usepackage{amssymb}
\usepackage{amsmath}
\usepackage{graphicx}
\usepackage{epstopdf}
\usepackage{bm}% bold math
\usepackage{gensymb}
\usepackage{soul}
\usepackage{sidecap}
\usepackage[normalem]{ulem}
\usepackage{times}

\usepackage{float}
\usepackage{enumerate}
\usepackage{multirow}
\usepackage{tabularx}
\usepackage{array}
\usepackage{url}
\usepackage{slantsc}
\usepackage{lmodern}
\usepackage[normalem]{ulem}
\usepackage{xparse}
\usepackage{soul}
\usepackage{xcolor}
\usepackage{graphicx}
\makeatletter
\NewDocumentCommand{\sotwo}{O{red}O{black}+m}
{%
	\begingroup
	\setulcolor{#1}%
	\setul{-.5ex}{.4pt}%
	\def\SOUL@uleverysyllable{%
		\rlap{%
			\color{#2}\the\SOUL@syllable
			\SOUL@setkern\SOUL@charkern}%
		\SOUL@ulunderline{%
			\phantom{\the\SOUL@syllable}}%
	}%
	\ul{#3}%
	\endgroup
}
\makeatother
\graphicspath{{figs/}}
\def\beq{\begin{equation}}
\def\eeq{\end{equation}}
\def\bea{\begin{eqnarray}}
\def\eea{\end{eqnarray}}

\begin{document}
\title{Fractional Nonlinear Schrodinger Equation Revisited}

\author {Morteza Nattagh Najafi}
\email{morteza.nattagh@gmail.com }
\affiliation{Department of Physics, University of Mohaghegh Ardabili, Ardabil, Iran}

\author {Fatemeh Foroughirad}
\affiliation{Department of Mathematics, University of Mohaghegh Ardabili, Ardabil, Iran}


\begin{abstract}
We investigate the space-time fractional nonlinear Schrödinger equation (FNLSE) incorporating the modified Riemann–Liouville derivative introduced by Jumari. The equation is characterized by two parameters: the fractional derivative parameter ($\alpha$, which captures the memory effects) and the non-linearity parameter ($a$). We present analytical solutions via three complementary approaches: the fractional Riccati method, the Adomian decomposition method, and the scaling method. The FNLSE is formulated in terms of generalized Hamiltonian and momentum operators, allowing a unified framework to explore various solution structures. A continuity equation is derived, and a general class of solutions based on Mittag-Leffler (ML) plane waves is proposed, from which generalized momentum and energy eigenvalues are systematically classified. Utilizing a generalized Wick rotation, we establish a connection between the FNLSE and a fractional Fokker–Planck equation governing a dual stochastic process, revealing links to $Q$-Gaussian statistics, where $Q=1-a$. Through separation of variables, we classify a family of solutions including chiral ML plane waves. Additionally, we construct Riccati-type bright and dark solitons and assess their stability using a dynamical distance metric. As $\alpha$ changes, i.e. the memory effects are tuned, the bright solitons transform to dark solitons, which is equivalent to focusing-defocusing transition. A series solution is developed via the Adomian decomposition technique, applied to both chiral and plane-wave cases. Finally, we reduce the dimensionality of the FNLSE using the scaling arguments, leading to self-similar solutions, and find pure-phase as well as Adomian-type solutions. Our results highlight the rich analytical landscape of the FNLSE and provide insights into its underlying physical and mathematical structure.
\end{abstract}

\maketitle
\tableofcontents 
%%%%%%%%%%%%%%%%%%%%%%%%%%%%%%%%%%%%%%%%%%%%%%%%%%%
\section{Introduction}\label{1}
The interplay between nonlinearity~\cite{thompson2002nonlinear,rega2020nonlinear,rhoads2010nonlinear,hilborn2000chaos} and memory effects~\cite{tarasov2017time,laskin2000fractional} lies at the heart of many complex quantum and classical phenomena. In particular, nonlinear extensions of the Schrödinger equation have proven indispensable for modeling a wide array of physical systems~\cite{pang2005quantum,abrams1998nonlinear,kaplan2022causal,kibble1978relativistic,polchinski1991weinberg,wallentowitz1997quantum}, ranging from Bose–Einstein condensates~\cite{rogel2013gross} to nonlinear optics~\cite{wallentowitz1997quantum,fibich2015nonlinear,he1999physics,longhi2015fractional} and classical and quantum plasmonics~\cite{enjieu2008nonlinear,papadopoulos2006non}. Combining nonlinearity with anomalous dispersion can lead to a rich variety of complex phenomena that go beyond what is seen in systems with either feature alone~\cite{metzler2000random}, like the emergence of non-Gaussian and heavy-tailed distributions~\cite{bhagavatula1993non,racz2021estimation,dos2018non}, modified scaling laws~\cite{castiglione1999strong}, long-range correlations and memory effects~\cite{wang1994correlation}, pattern formation and instabilities \cite{colombo2012nonlinear,muller2002morphological}, modified transport and relaxation dynamics~\cite{goychuk2021nonequilibrium}, and multifractality and complex scaling structures~\cite{seckler2024multifractal}. This approach effectively captures the non-local and long-range temporal correlations inherent in nonlinear quantum systems with anomalous dispersion—such as those described by the fractional nonlinear Schrödinger equation (FNLSE)~\cite{tarasov2005fractional}—as well as in nonlinear anomalous diffusion processes, where the mean squared displacement scales as $t^\alpha$, with \( \alpha > 0 \) a behavior naturally modeled using the tools of fractional calculus~\cite{shlesinger1987levy,west1997fractional,metzler2000random}.

A path integral formulation using Lévy flights was suggested by Laskin leading to a fractional version of the Schrödinger equation~\cite{laskin2002fractional}, which laid the foundation for subsequent exploration of nonlinear versions (FNLSE). In this case, due to the long-range interactions and nonlocal effects~\cite{laskin2006nonlinear}, the quantum propagators transition from conventional Gaussian profiles to heavy-tailed, power-law decays. Various aspects of FNLSE have been addressed, including the Markov property and the Wick rotation~\cite{naber2004time}, the conservation of the probability measures and the particles' absorption in the potentials~\cite{dong2008space}, integrability~\cite{zakharov1974complete,fokas2016integrable,gerdjikov2017complete,lenells2008novel,ablowitz2016inverse}, the scattering theory~\cite{ginibre1979class,nakanishi2002remarks,ozawa1991long}, self-similar solutions~\cite{kruglov2003exact,chen2005chirped,lin2006solitary,marikhin2000self,budd1999new,budd2006computation,lin2006solitary,perez2004self}, and the soliton solutions~\cite{serkin2000novel,gurses2018nonlocal,ali2017soliton,feng2018general,chowdury2014soliton,liaqat2022novel,carter1987squeezing,gu2013soliton,tarasov2005fractional,ablowitz2022fractional,ahmad2023soliton,ahmad2023novel,chen2018optical,pavlyshynets2025solitons,liu2025experimental}. 

Numerous analytical and semi-analytical methods have been developed to obtain soliton solutions of FNLSE, including both classical and fractional forms. Early studies on soliton solutions of the damped NLSE in its Fourier-transformed form were initiated by Pereira et al.~\cite{pereira1977nonliner}, and a more comprehensive treatment of the appropriate damping conditions for exact soliton solutions was later provided in~\cite{pereira1977soliton}. These include the Bäcklund transformation~\cite{grecu1984buacklund}, the Hirota bilinear method~\cite{liu2008soliton}, and several other integrability techniques. With the growing interest in fractional differential equations, numerous techniques have also been adapted or developed to handle fractional-order nonlinear problems. These include the sub-equation method~\cite{bekir2015exact,zhang2011fractional,guo2012improved}, the \(G'/G\) expansion method~\cite{zheng2012g,wang2008g}, the fractional Bäcklund transformation~\cite{lu2012backlund}, and powerful semi-analytical approaches such as the Adomian decomposition method~\cite{el2006adomian,el2010adomian,daftardar2005adomian}, the variational iteration method~\cite{wu2010fractional,guo2011fractional}, and the homotopy perturbation method~\cite{ganji2008application}. Recently, a fractional Riccati expansion method was introduced to derive exact analytical soliton solutions based on fractional trigonometric and hyperbolic functions~\cite{abdel2016analytical,abdel2013solution}, providing a promising framework for addressing nonlinear fractional equations, including FNLSE. To date, there has been no systematic investigation of a self-similar solutions for FNLSE, leaving a gap that invites further exploration. 

Building on the foundations laid by previous studies, the present work focuses on the analytical investigation of the space-time FNLSE incorporating the modified Riemann–Liouville derivative introduced by Jumari. This generalized formulation extends the classical NLSE by incorporating nonlocality and memory effects through fractional differentiation in both space and time. We formulate a fractional continuity equation, helping us to design a general solution ansatz based on ML exponential functions, and also a classical velocity derived from the ML phase. By applying a generalized Wick rotation, we uncover a duality between the FNLSE and a fractional Fokker–Planck equation, describing a stochastic process with non-Gaussian behavior characterized by $Q$-Gaussian distributions in some limit, where $Q=1-a$. The central aim of this paper is to derive and classify analytical solutions of the FNLSE using three distinct, yet complementary, methods: the fractional Riccati Method, the Adomian decomposition method, and the scaling method. Each of these approaches offers a unique perspective and mathematical structure for handling the inherent nonlocality and nonlinearity of the equation. Through this multifaceted analytical lens, we construct a rich spectrum of solution types, including Mittag-Leffler (ML) plane waves, chiral modes, bright and dark solitons, and self-similar solutions.

By combining the Riccati method with a stability analysis based on a dynamical distance metric, we also provide new insights into the robustness of fractional solitons. The Adomian decomposition approach allows for the systematic construction of series solutions, especially valuable in regimes where exact closed-form expressions are inaccessible. Finally, through the scaling method, we explore self-similar behaviors and scale-invariant dynamics, which may be relevant in both theoretical and experimental scenarios involving critical phenomena or scale-free media.

\section{Aims and Findings of This Study}
After presenting the general methods—namely the Riccati, Adomian, and scaling techniques—and defining key concepts in the following section, we turn our attention to the central equation of this study, Eq.~\ref{Eq:NLSE}, i.e. the fractional nonlinear Schrödinger equation (FNLSE) expressed in terms of a generalized Hamiltonian and momentum (Eq.~\ref{Eq:GHM}). The main findings of the present paper are itemized as follows:

\begin{itemize}
    \item For FNLSE we derive a continuity equation (Eq.~\ref{continuity}). By employing the identity in Eq.~\ref{Eq:MittagSol}, we introduce a general form based on the Mittag-Leffler (ML) plane wave, leading to a proposed general structure for the solutions (Eq.~\ref{Eq:decomposition}). 
    \item In particular, the classical velocity is obtained from Eq.~\ref{Eq:velocity}, which is subsequently used in the formulation of the generalized Hamilton-Jacobi equation (Eq.~\ref{Eq:HamiltonJacobi}); the corresponding expression for the classical energy is provided in Eq.~\ref{Eq:classicalEnergy}. 
    \item The ML plane wave itself is defined in Eq.~\ref{Eq:MLPW}, from which the eigenvalues of the generalized momentum and Hamiltonian are systematically classified (Eqs.~\ref{Eq:momentum} and \ref{Eq:Hamiltonian}, respectively). 
    \item Through a generalized Wick rotation, we propose an interpretation in terms of a generalized Fokker-Planck equation describing a dual stochastic process (Sec.~\ref{SEC:Wicksec}), where we also explore its connection to $Q$-Gaussian distributions (Eq.~\ref{Eq:QGaussian-Main}).
    \item By applying the method of separation of variables (Eq.~\ref{Eq.17}), we classify the solutions of the FNLSE. In particular, a chiral ML plane wave is examined in Sec.~\ref{SEC:MLPW}. 
    \item A family of Riccati-type solitons (Riccati solitons for short) is proposed and analyzed in Sec.~\ref{SEC:RiccatiMain}; these solutions are summarized in Table~\ref{tab:solutions} and illustrated in Figs.~\ref{fig:Solitorysol2},~\ref{fig:Solitorysol3}, and~\ref{fig:Solitorysol1}. The stability of these solitonic solutions is another important issue, which is addressed in Figs.~\ref{fig:stability1}, \ref{fig:stability2}, and~\ref{fig:stability3}, using a measure of state distance defined in Eq.~\ref{Eq:stabilityMain}. 
    \item Additionally, a series expansion of the solutions is developed in Sec.~~\ref{SEC:Adomian} using the Adomian decomposition method, applied to both the chiral case (Sec.~\ref{SEC:chiral}) and the plane-wave case (Sec.~\ref{SEC:PWAdomian}). The explicit series expansions are given in Eqs.~\ref{Eq:Adomian1} and \ref{Eq:Adomian2}.
    \item Self-similar solutions are investigated in Sec.~\ref{SEC:SelfSimilar}, based on the scaling relation given in Eq.~\ref{Eq:scalingMain}, which leads to the formulation in Eq.~\ref{eq97}. In addition to a purely phase-based solution derived from the ML phases (Eq.~\ref{SEC:ScalingAdomian}), an Adomian-type solution (Adomian solution for short) is also obtained and analyzed in Sec.~\ref{Eq.scalingAdomian}, which leads to the series expansion Eq.~\ref{Eq.scalingAdomian}.
\end{itemize}

\section{Definitions and Methods}\label{SEC:Methods}
This section is devoted to identifying the notations, definitions and methods employed to handle the partial differential equations. In the following section we introduce the modified Riemann-Liouville derivative based on which the non-linear Schrodinger equation is expressed, and also the methods we employed to find the solutions.

\subsection{Modified Riemann-Liouvulle derivative}
Consider a $1+1$ general fractional nonlinear differential equation, described by the following equation
\begin{equation}
\mathcal{P}(u, \mathcal{D}_t^{\alpha} u, \mathcal{D}_x^{\beta} u, \dots) = 0, \label{polynimial}
\end{equation}
where \( \mathcal{P} \) represents a polynomial in \( u \) and its partial derivatives. This equation can be a non-linear Schrodinger equation~\cite{gutkin1988quantum}, or any Fokker-Planck-type or dynamical equation~\cite{feireisl2016mathematical}. In this equation \( \mathcal{D}_{\zeta}^{\alpha} u \) ($\zeta=x,t$) denotes the the modified (Jumarie) Riemann-Liouville (RL) derivative defined as follows for every function $u(\zeta)$:
 \begin{equation}
\mathcal{D}_{\zeta}^\alpha u(\zeta) =
\begin{cases}
\frac{1}{\Gamma(-\alpha)} \int_0^\infty (\zeta-\zeta')^{-\alpha-1}(u(\zeta') - u(0)) \, d\zeta', \\[10pt]
\frac{1}{\Gamma(1-\alpha)} \frac{d}{d\zeta} \int_0^\infty (\zeta-\zeta')^{-\alpha}(u(\zeta') - u(0)) \, d\zeta', \\[10pt]
\frac{1}{\Gamma(n+1-\alpha)} \frac{d^n}{d\zeta^n} \int_0^\infty (\zeta-\zeta')^{n-\alpha}(u(\zeta') - u(0)) \, d\zeta',
\end{cases}
\end{equation}   
where $\alpha$ is an external (fractional) factor, being $\alpha<0$, $0<\alpha<1$, and $n\le \alpha<n+1$ ($n\ge 1$) for the first, second and third lines respectively. $\Gamma(\cdot)$ denotes the gamma function. Some properties of this operator are as follows:
\begin{itemize}
    \item The Modified RL-derivative of a constant is zero.
    \item It preserves the Leibniz rule.
    \item  It maintains the fractional chain rule for composite functions.
    
\end{itemize}
 The inverse of the Jumarie derivative is the fractional RL integral, defined as: 
\begin{equation}
\begin{split}
   \mathcal{J}^\alpha_{0} f(x) &\equiv  \frac{1}{\Gamma(\alpha+1)}\int_{0}^xf(\zeta)\text{d}^\alpha \zeta
   \\&\equiv\frac{1}{\Gamma(\alpha)} 
    \int_{0}^{x} (x-\zeta)^{\alpha-1} f(\zeta) \, \text{d}\zeta,
    \label{Eq:RLIntegral}
\end{split}
\end{equation}
is the RL integral, known also as the inverse of the Jumarie derivative:
\begin{equation}
\mathcal{J}^\alpha_0\mathcal{D}_x^{\alpha}f(x)=f(x)-f(0),
\end{equation}
so that it is often represented by the inverse of the Jumaries derivative $\mathcal{J}_x^{\alpha}\equiv \left(\mathcal{D}_x^{\alpha}\right)^{-1}$. Some other properties, of Jumarie derivatives are presented in ~\ref{App:RL}, which are going to be employed in the remaining of this paper.

\subsection{Fractional Riccati Method}\label{SEC:RicattiMain}
The fractional (modified RL derivative) Riccati method was introduced in ~\cite{abdel2013solution}, which can be utilized to analyze various nonlinear systems ~\cite{abdel2016analytical,altwaty2021optical,wang2020vector,lu2020fractional}. In this section we explore briefly the key steps of fractional Riccati Method to find a solution of the Eq.~\ref{polynimial}. These steps are outlined in the following:
\begin{itemize}
    \item \textbf{Step 1.} We reduce the dimensionality of the system. A very common method to this end is to apply the traveling wave transformation as follows:
\begin{equation}
u(x, t) = u(\xi), \quad \xi = kx + \omega t, \quad 0 < \alpha < 1, 
\end{equation}
where \( k \) and \( \omega \) are constants to be determined later. We call these solutions as the \textit{chiral solutions}, for which the current is chiral. It is notable that there are other methods to reduce the dimensionality, like the scaling arguments. The Eq.~\eqref{polynimial} is then reduced to the following nonlinear fractional ordinary differential equation (FODE) for \( u = u(\xi) \),
\begin{equation}
\tilde{\mathcal{P}}(u, \omega^\alpha \mathcal{D}_\xi^\alpha u, k^\beta \mathcal{D}_\xi^\beta u, \dots) = 0.\label{polynimial2} 
\end{equation}
\item \textbf{Step 2.} The following trial finite power series solution \( u(\xi) \) is considered
\begin{equation}
 u(\xi) = a_0 + \sum_{i=1}^{N} a_i F^i(\xi), \quad a_N \neq 0,\label{expansion} 
\end{equation}
where $a_i$'s $i = 0, 1, 2, \dots, N)$ are constants to be determined later, \( N \) represents a positive integer determined by balancing the highest-order linear term with the nonlinear term in Eq. \eqref{polynimial2}. In this equation \( F = F(\xi) \) is a general function that fulfills the following fractional Riccati equation
\begin{equation}
 \mathcal{D}_\xi^\alpha F = A + B F^2, \quad 0 < \alpha \leq 1,
 \label{Eq:Fequation}
\end{equation}
where $A$ and $B$ are constants. A list of possible choices for $F$ based on different values of $A$ and $B$ is provided in Appendix \ref{SEC:Ricatti}.
\item \textbf{Step 3.} 
To find closed-form solutions, we need to truncate the series expansion of $u$ in terms of $F$ (Eq.~\ref{expansion}). For this truncation we obey a balance process as follows: To determine the balance term \( N \), we focus on the highest exponents of \( u \) and the highest-order derivative present in the equation. Specifically, we equate their associated balance terms by considering the nonlinear term \( u^{a} \) with balance term \( aN \), and the highest-order derivative term, such as \( \mathcal{D}^{2\alpha}u \), with balance term \( N+2 \). Substituting the fractional Riccati expansion, Eq. \eqref{expansion}, into Eq. \eqref{polynimial2}, and using Eq  \eqref{Eq:Fequation}, transforms the left-hand side of Eq. \eqref{polynimial2} into a polynomial in \( F(\xi) \). 
\end{itemize}
Then, by setting each coefficient of this polynomial to zero, a system of algebraic equations is obtained. Solving this system allows the values of \( a_0, a_1, \dots, a_N, k, \) and \( \omega \) to be expressed in terms of the parameters $A$ and $B$.

\subsection{Adomian Decomposition Method}
The Adomian decomposition method (ADM), initially introduced and developed by George Adomian in \cite{adomian1988review}. In this approach, Eq. \eqref{polynimial} is expressed in terms of its linear and nonlinear components as follows:

\begin{equation}
\mathcal{L}[u] + \mathcal{R}[u] + \mathcal{N}[u] = 0,
\end{equation}
where \( \mathcal{L} \) and \( \mathcal{R} \) are linear operators containing fractional Jumarie RL derivatives, and \( \mathcal{N} \) represents a nonlinear operator. To simplify the equation, mostly the inverse operator of the lowest order fractional derivative, \( \mathcal{L}^{-1} \), is applied to both sides, resulting in

\begin{equation}
u =f -\mathcal{L}^{-1}[\mathcal{R}(u)] - \mathcal{L}^{-1}[\mathcal{N}(u)],\label{ado}
\end{equation}
where the function \( f \) represents the term that arises from applying the inverse of \( \mathcal{L}[u] \), which is typically provided in the given conditions.~In this method, $u$ and $\mathcal{N}(u)$ are represented as series expansions, as discussed in Appendix \ref{SEC:Adomian}. Substituting these expansions into Eq.~\eqref{ado}, the following recurrence relations are derived:
\begin{equation}
u_0 =f,
\end{equation}
\begin{equation}
u_{n+1} = -\mathcal{L}^{-1}[\mathcal{R}(u_n)] - \mathcal{L}^{-1}[A_n].
\end{equation}
Having determined these components, we obtain a  solution for $u$ in series form.
\subsection{Scaling Method}
 Scaling laws provide deep insights into phenomena such as self-similarity. The term "self-similarity" refers to solutions at a specific time \( t_1 \) that closely resemble those at an earlier time \( t_0 \) \cite{barenblatt1996scaling}. To derive "self-similar" solutions for Eq.~\eqref{polynimial}, following scaling transformations are applied:

\begin{equation}
\begin{split}
& t \rightarrow \lambda t, \quad x \rightarrow \lambda^{\beta'} x, \quad u \rightarrow \lambda^{-\eta} u,\\
&\mathcal{D}^\alpha_t \rightarrow \lambda^{-\alpha} \mathcal{D}^\alpha_t, \quad \mathcal{D}^{\beta}_x \rightarrow \lambda^{-\beta\beta'} \mathcal{D}^{\beta}_x,\label{transformation}
 \end{split}   
\end{equation}
where~$\lambda$ is scaling factor. This scaling transformation results in a new polynomial, namely $P\to P'$, where
\begin{equation}
\mathcal{P}'(u, \mathcal{D}^\alpha_t u, \mathcal{D}^{\beta}_x u, \dots)\equiv \mathcal{P}(\lambda^{-\eta} u, \lambda^{-\alpha'} \mathcal{D}^\alpha_t u,\lambda^{-\beta'' } \mathcal{D}^{\beta}_x u, \dots).
\end{equation}
where $\alpha'\equiv \alpha+\eta$ and $\beta''\equiv \beta\beta'+\eta$. Let's suppose that $\mathcal{P}'$ is a general transformation of $\mathcal{P}$ described by $f(\mathcal{P},\lambda)$, so that
\begin{equation}
 \mathcal{P}'= f( \mathcal{P}(u, \mathcal{D}^\alpha_t u, \mathcal{D}^{\beta}_x u, \dots),\lambda)
\end{equation}
where $f(y,\lambda)$ is a general transformation which is dictated by the governing equation. Then the system under study has scale invariance if a solution of $f( y,\lambda)=0$ is $y=0$ which corresponds to the original equation, i.e.
\begin{equation}\label{Eq:govEq}
\mathcal{P}(u, \mathcal{D}^\alpha_t u, \mathcal{D}^{\beta}_x u, \dots)=0.
\end{equation}
An important example of a self-similar solution is given by \( f(y, \lambda) = \lambda^{\xi} y^{\xi'} \), where \( \xi \) and $\xi'$ are some exponents. Then one trivially finds that the Eq.~\eqref{Eq:govEq} is concluded, telling us that the system is scale invariant. One may reduce the number of independent variables of the problem using the scaling arguments.

\section{Fractional Non-linear Schrodinger Equation}
The FNLSE is expressed as:
\begin{equation}  
    i \mathcal{D}_t^{\alpha} u - p \mathcal{D}_x^{\beta} u - q |u|^a u = 0. 
    \label{Eq:NLSE}  
\end{equation}
where \( p \) and \( q \) are real constants, and \( a \) is a positive parameter, and $\alpha$ and $\beta$ are fractionalization parameters, known as the fractional orders of FNLSE. While $\beta$ can be any arbitrary exponent, to avoid unnecessary complications we consider the case 
\begin{equation}
\beta=2\alpha.
\label{Eq:betaalpha}
\end{equation}
In some sections (like the scaling solutions) we consider the $\beta$ as a general independent exponent to maintain the formalism as general as possible, but in the end we apply the Eq.~\eqref{Eq:betaalpha}. The ordinary Schrodinger equation is retrieved by taking the limit $\beta=2\alpha=2$ and $a=0$.\\

The FNLSE can also be written in the following form:
\begin{equation}
\hat{H}^{(\alpha)}u=-p\left[\hat{p}^{(\alpha)}\right]^2u+q|u|^au.
\end{equation}
where 
\begin{equation}
\hat{H}^{(\alpha)}\equiv i\mathcal{D}_t^\alpha,\ \hat{p}^{(\alpha)}\equiv-i\mathcal{D}^\alpha_x,
\label{Eq:GHM}
\end{equation}
are the generalized Hamiltonian and generalized momentum respectively. If we multiply Eq.~\eqref{Eq:NLSE} by the conjugate of the respective solution,$u^*$, and multiply the conjugate of Eq.~\eqref{Eq:NLSE} by $u$, and subtract the results, we find 
\begin{equation}
    \left( u^* \mathcal{D}_t^{\alpha} u + u \mathcal{D}_t^{\alpha} u^* \right) 
    + i p \left( u^* \mathcal{D}_x^{2\alpha} u - u \mathcal{D}_x^{2\alpha} u^* \right) = 0.
\end{equation}
Then we define the particle density and current respectively as
\begin{equation}
    \rho^{(\alpha)}(x,t) = u^* u, \quad 
    j^{(\alpha)}(x,t) = i p \left( u^* \mathcal{D}^\alpha_x u - u \mathcal{D}^\alpha_x u^* \right).\label{current}
\end{equation}
 
then the continuity equation reads
\begin{equation}
    \mathcal{D}^\alpha_t \rho^{(\alpha)} + \mathcal{D}^\alpha_x j^{(\alpha)} = 0.\label{continuity}
\end{equation}
The physical interpretation of this equation becomes clear if we take a fractional integral around $dx^\alpha $, and using ~\eqref{A7} and \eqref{A8} , resulting to the following formula
\begin{equation}
    \mathcal{D}^\alpha_t m^{(\alpha)}=j^{(\alpha)}(x^-,t)-j^{(\alpha)}(x^+,t)
    \label{Eq:massConserv}
\end{equation}
where $m^{(\alpha)}\equiv\mathcal{J}^\alpha_x \rho^{(\alpha)}(x,t)$ is the total mass, $x^+$ ($x^-$) is the right (left) boundary point, obtained using the RL integral Eq.~\eqref{Eq:RLIntegral}. The Eq.~\eqref{Eq:massConserv} indicates that the total fractional current flowing in and out is not only determined by the instantaneous change in fractional particle density but also by its entire past history.

\subsection{Mittag-Leffler Plane Waves and Stationary States} 
An important identity that helps to find solutions for the differential equation that include Jumarie derivative is one-parameter Mittag-Leffler function, which is defined as
\begin{equation}
E_\alpha(x) = \sum_{k=0}^\infty \frac{x^{\alpha k}}{\Gamma(\alpha k + 1)}
\end{equation}
This function has a vast application in fractional non-linear systems, for a good review see~\cite{haubold2011mittag}. In the integer limit, we have
\begin{equation}
\lim_{\alpha\to 1}E_\alpha(x)=e^x.
\end{equation}
It is helpful to define a Mittag-Leffler exponential as follows
\begin{equation}
\mathcal{E}_{\alpha}(x)\equiv \sqrt{\frac{E_{\alpha}\left(ix^\alpha\right)}{E_{\alpha}\left(-ix^\alpha\right)}}, \ x\ge 0.
\label{Eq:MLphase}
\end{equation}
Note that 
\begin{equation}
\lim_{\alpha\to 1}\mathcal{E}_{\alpha}(x)=\exp\left[ix\right].
\end{equation}
In the following we will see the importance of this function. Firstly, utilizing equations~\eqref{eq:A.5} and~\eqref{Mittag}, one can easily verify that $\mathcal{E}_{\alpha}(x)$ is an eigenfunction of the generalized momentum operator $\hat{p}^{(\alpha)}$ (Eq.~\ref{Eq:GHM}) with an eigenvalue $+1$:
\begin{equation}
\hat{p}^{(\alpha)}\mathcal{E}_{\alpha}(x)=-i\mathcal{D}^\alpha_x \mathcal{E}_{\alpha}(x) = \mathcal{E}_{\alpha}(x).
\label{Eq:MittagSol}
\end{equation}
A more direct interpretation of FNLSE is obtained by taking the standard representation of the solutions as 
\begin{equation}
    u(x,t) = \sqrt{\rho^{(\alpha)}(x,t)} e^{iS_g^{(\alpha)}(x,t)},
    \label{Eq:decomposition}
\end{equation}
where \( S^{(\alpha)}_g(x,t) \) is a real phase function. We represent this phase in terms of the ML exponential as follows:
\begin{equation}
    e^{iS_g^{(\alpha)}(x,t)} \equiv \mathcal{E}_{\alpha}(g(x,t)).
\label{Eq:phase}
\end{equation}
Here $g(x,t)$ is a real \textit{non-negative} function that is related to the current density of the system. The positivity of $g(x,t)$ guarantees that the phase $S_g^{\alpha}$ is real. By incorporating Eq.~\eqref{Eq:phase} into Eq.~\eqref{current}, and using \eqref{eq:A6}, the current density is found to be
\begin{equation}
    j^{(\alpha)}(x,t) = -2p \rho^{(\alpha)}(x,t) \left( \partial_xg(x,t) \right)^\alpha.
    \label{Eq:velocity}
\end{equation}
This tells us that the velocity is given by
\begin{equation}
\begin{split}
 v^{(\alpha)}(x,t)&\equiv \frac{j^{(\alpha)}(x,t)}{\rho^{(\alpha)}(x,t)}=-2p\left( \partial_xg(x,t) \right)^\alpha\\
 &=-\frac{2p}{\Gamma(\alpha+1)}\mathcal{D}_x^{\alpha}\left(g(x,t)^\alpha\right)\\
 &=-\frac{2ip}{\Gamma(\alpha+1)}\hat{p}^{(\alpha)}g(x,t)^\alpha
\end{split}
\label{Eq:velocity}
\end{equation}
where $\partial_x\equiv \mathcal{D}_x^1$ is an ordinary partial derivative. This is the generalized classical velocity of particles in terms of the generalized gradient of $g(x,t)$. We see that in the $\alpha\to 1$ one retrieves ordinary relation for the velocity.\\

The ML \textit{plane wave} is subsequently defined as
\begin{equation}
w^{(\alpha)}_k(x)\equiv A_w\mathcal{E}_\alpha(k^{1/\alpha}|x|),
\label{Eq:MLPW}
\end{equation}
where $|.|$ is an absolute value, and $A_w$ is a normalization constant. This function, which goes to a simple plane wave in the $\alpha\to 1$ limit, has a singularity at $x=0$ as a result of the absolute value. It is an eigenvector of the generalized momentum operator:
\begin{equation}
\hat{p}^{(\alpha)}w^{(\alpha)}_k(x)=k w^{(\alpha)}_k(x).
\label{Eq:momentum}
\end{equation}
Also, working in the time domain, one easily shows that the generalized Hamiltonian $\hat{H}^{(\alpha)}$ (Eq.~\eqref{Eq:GHM}) has the eigenvectors:
\begin{equation}
\hat{H}^{(\alpha)}\phi^{(\alpha)}_E(t)=E \phi^{(\alpha)}_E(t),
\label{Eq:Hamiltonian}
\end{equation}
where
\begin{equation}
\phi^{(\alpha)}_E(t)\equiv \mathcal{E}_{\alpha}\left((-E)^{1/\alpha}|t|\right).
\label{Eq:stationary}
\end{equation}
These identities are used in the following section for finding the stationary current of quantum particles described by FNLSE.
Another real-valued function is the ML Gaussian function 
\begin{equation}
G_{\alpha}\left[\zeta,x\right]\equiv e_{\alpha}(\zeta x^2),
\label{Eq:GaussianMain}
\end{equation}
where
\begin{equation}
e_{\alpha}(y)\equiv \sqrt{\frac{E_{\alpha}\left(y^\alpha\right)}{E_{\alpha}\left(-y^\alpha\right)}}, \ y\ge 0. \label{ealpha}
\end{equation}
Note that
\begin{equation}
e_{\alpha}(y)=\mathcal{E}_{\alpha}\left((-i)^{1/\alpha}y\right),
\end{equation}
is a real function for $x\ge 0$, telling us that
\begin{equation}
\begin{split}
\mathcal{D}_y^{\alpha}e_\alpha(y)&=-i\mathcal{D}_\xi^\alpha\mathcal{E}_{\alpha}(\xi)|_{\xi=(-i)^{1/\alpha}y}\\
&=\mathcal{E}_\alpha\left((-i)^{1/\alpha}y\right)=e_\alpha(y),
\end{split}
\end{equation}
i.e. it is the exact eigenvector of $\mathcal{D}_y^\alpha$ with an eigenvalue $+1$. Note that $G_{\alpha}\left[\zeta,x\right]$ tends to the ordinary Gaussian form $\exp(-\zeta x^2)$ when $\alpha\to 1$.

\subsection{The generalized Hamilton-Jacobi equation}
Here we write the FNLSE using the decomposition Eq.~\ref{Eq:decomposition}, and the Eq.~\ref{eq.A4}, resulting to
\begin{widetext}
\begin{equation}
i\mathcal{D}_t^\alpha \sqrt{\rho^{(\alpha)}}-\sqrt{\rho^{(\alpha)}}\left(\partial_tg\right)^\alpha -p\mathcal{D}_x^{2\alpha}\sqrt{\rho^{(\alpha)}}+p\sqrt{\rho^{(\alpha)}}\left(\partial_xg\right)^{2\alpha}-2ip\left(\partial_xg\right)^\alpha\mathcal{D}_x^\alpha\sqrt{\rho^{(\alpha)}}-ip\sqrt{\rho^{(\alpha)}}\mathcal{D}_x\left(\partial_xg\right)^\alpha-q(\rho^{(\alpha)})^{\frac{a+1}{2}}=0.
\end{equation}
\end{widetext}
An important limit is the ``slow-varying function" approximation. To this end, we consider the normal limit of the parameters, defined as the parameters which produce the ordinary Schrodinger equation in the limit $\alpha\to 1$ and $a\to 0$, which tells us that $p$ corresponds to $-\frac{\hbar}{2m}$, and $g\to \frac{S}{\hbar}$, where $S$ is the ordinary phase of the wave function. Keeping everything to the leading order as $\hbar\to 0$ gives
\begin{equation}
\left(\partial_tg\right)^\alpha-p\left(\partial_xg\right)^{2\alpha} +q(\rho^{(\alpha)})^{\frac{a}{2}}=0,
\label{Eq:HamiltonJacobi}
\end{equation}
which is a generalized Hamilton-Jacobi equation in the classical mechanics. We can further simplify the equations by considering the stationary state, defined using the equation
\begin{equation}
g(x,t)\equiv W(x)+(-E)^{\frac{1}{\alpha}}t,
\end{equation}
where $W(x)$ is Hamilton's characteristic function and $E$ is a constant (to understand the reason of the $1/\alpha$ exponent for $E$, refer to Eq.~\eqref{Eq:stationary}). This gives
\begin{equation}
-E-p(W')^{2\alpha}+V(\rho^{(\alpha)})=0.
\end{equation}
where $V(\rho^{(\alpha)})\equiv q(\rho^{(\alpha)})^{\frac{a}{2}}$. Note that, according to Eq.~\eqref{Eq:velocity}, $(W')^\alpha=-\frac{v^{(\alpha)}}{2p}$, so that
\begin{equation}
E=-\frac{(v^{(\alpha)})^2}{4p}+V(\rho^{(\alpha)}).
\label{Eq:classicalEnergy}
\end{equation}
This is reminiscent of the classical energy of particles. This classical representation, also known as the Thomas--Fermi approximation, has important applications, including local energy estimation in the local density approximation (LDA) within condensed matter systems~\cite{????}.
, and the physics of dark matter~\cite{hochberg2016detecting}.

\subsection{A Solution for the Case $a=0$}\label{subsection:c}
Before going into more details of the techniques to solve FNLSE, we present a simple type of solution found for $a=0$ and general $\alpha$ using the method of separation of variables. To this end 
we consider the solution as
 \begin{equation}
     u_{a=0}(x,t)= X(x) \phi_E(t).\label{separable}
 \end{equation}
which implies 
\begin{equation}
\frac{i}{\phi_E} \mathcal{D}^\alpha_t \phi_E- \frac{p}{X} \mathcal{D}^{2\alpha}_x X=q.
\end{equation}
In this equation the sub-script $E$ shows the generalized \textit{energy} of the system, defined in Eq.~\ref{Eq:Hamiltonian}. Using the fact that the first (second) term is a pure function of $t$ ($x$) one finds
\begin{equation}
  i\mathcal{D}^{\alpha}_t \phi_E=E\phi_E,\ 
  \mathcal{D}^{2\alpha}_x X=-\frac{k}{p}X,
\end{equation}
where $E$ and $k$ are real numbers satisfying $E+k=q$. We choose $k$ so that $\frac{k}{p}>0$. Using the solution found in SEC.~\ref{AppE}, one finds
\begin{equation}
u_{a=0}(x,t)=u_0e^{-\frac{k}{p}x^{2\alpha}}e^{-iEt^\alpha},
\end{equation}
where $u_0$ is a constant. $\phi_E$ is reminiscent of stationary solutions for the ordinary Schrodinger equation, which includes an energy-dependent phase.
\subsection{Separable Variable Solutions}
Important classes of FNLSE are found using the separation of variables. We, first, apply the following transformation
\begin{equation}
    u(x,t) = u_1(\xi) u_2(\eta), \quad \xi = kx + \omega t, \quad \eta = rx + st,
    \label{Eq.17}
\end{equation}
where ~$k$,~$\omega$,~$r$,~$s$ are constants to be determined later.
Substituting Eq.~\eqref{Eq.17} into Eq.~\eqref{Eq:NLSE} and using \eqref{eq:A6} as well as 
 \eqref{eq.A4}, we obtain
\begin{equation}
\begin{split}
  & -p \left[k^{2\alpha} u_2 \mathcal{D}^{2\alpha}_\xi u_1 + r^{2\alpha} u_1 \mathcal{D}^{2\alpha}_\eta u_2 + 2 k^\alpha r^\alpha \mathcal{D}^\alpha_\xi u_1 \mathcal{D}^\alpha_\eta u_2 \right] \\
  & + i \left[\omega^\alpha u_2 \mathcal{D}^\alpha_\xi u_1 + s^\alpha u_1 \mathcal{D}^\alpha_\eta u_2 \right]  - q |u_1|^a |u_2|^a u_1 u_2 = 0.
\end{split}
\label{Eq.18}
\end{equation}

In the following sections, we present different classes of solutions for \( u \), considering various cases for \( u_2 \) and subsequently finding the corresponding solutions for \( u_1 \).
\subsection{Generalized Wick's Rotation: Stochastic Phenomena Interpretation}\label{SEC:Wicksec}
In the ordinary systems, the Wick's rotation includes $\frac{\pi}{2}$ rotation in the complex time plane. For our system, using the following generalized Wick's rotation:
\begin{equation}
t\to \tau \equiv i^\alpha t=e^{\frac{i\alpha\pi}{2}}t,
\label{Eq:WickRotation}
\end{equation}
involving $\frac{\alpha\pi}{2}$ rotation in the complex time plane, one transforms the fNLSE to a Fokker-Planck equation of a stochastic process. More precisely, using Eq.~\ref{Eq:NLSE} one finds
\begin{equation}
\mathcal{D}_\tau^\alpha P(x,\tau)=p\mathcal{D}_x^{\beta} P(x,\tau)+qP(x,\tau)^{a+1}. \label{Wick}
\end{equation}
where $P(x,\tau)$ is a real-valued probability density of a stochastic process $x(\tau)$. Here, we consider $\beta=2 \alpha$. To solve Eq.\eqref{Wick}, as provided in appendix \ref{AppD} , for the trial function $P(x,\tau)=P_0(x,\tau)Q(x,\tau)$, where $P_0$ is an auxiliary function that satisfies 
\begin{equation}
\mathcal{D}_\tau^\alpha P_0(x,\tau)=qQ(x,\tau)^aP_0(x,\tau)^{a+1}.
\label{Eq:main-P0}
\end{equation}
Then the Eq.~\ref{Wick} casts to
\begin{equation}
\mathcal{D}_\tau^{\alpha}Q(x,\tau)=\frac{q}{P_0(x,\tau)}\mathcal{D}_x^{2\alpha}\left(P_0(x,\tau)Q(x,\tau)\right), \label{Eq:FPE}
\end{equation}
which is a generalized Fokker-Planck equation for $Q$. In this equation, $P_0(x,\tau)$ acts as a stochastic force in an overdamped dynamical system. While the equation appears linear in $Q(x,\tau)$, its coupling to Eq.~\ref{Eq:main-P0} reveals the underlying nonlinearity.\\

\textit{For $\alpha=1$} the solution is found to be
\begin{equation}
P_0(x,\tau)^{-a}=P_0(x,\tau_0)^{-a}-qa\int_{\tau_0}^{\tau}Q(x,\tau')^a\text{d}\tau'.
\label{Eq:P0int-main}
\end{equation}
One may try the scaling relation 
\begin{equation}
Q(x,\tau)\equiv\tau^{-\gamma}F\left(\frac{x}{\tau^{\gamma}}\right) \label{D6}
\end{equation}
to evaluate the integral Eq.~\ref{Eq:P0int-main}, where $F(y)$ is a slow-varying function, and $\gamma$ is an exponent related to $\alpha$ and $a$. Defining $\zeta\equiv \frac{x}{\tau^\gamma}$, so that $\text{d}\tau=-\frac{x^\gamma}{\gamma}\zeta^{-(\gamma+1)}\text{d}\zeta$, one finds (Eq.~\ref{D6})
\begin{equation}
P_0(x,\tau)\approx \frac{P_0(x,\tau_0)}{\mathbb{E}_{Q}\left(-C(\tau_0^{-\eta}-\tau^{-\eta})\right)},
\label{Eq:QGaussian-Main}
\end{equation}
where $\tau_0$ is a reference time, and $Q=1-a$ and
\begin{equation}
C\equiv \frac{q\bar{F}^a}{\gamma(a-\gamma)P_0(x,\tau_0)^a}, \ \eta\equiv \gamma(a-\gamma).
\end{equation}
In Eq.~\ref{Eq:QGaussian-Main}
\begin{equation}
\mathbb{E}_{Q}(y)\equiv \left(1+(1-Q)y\right)^{\frac{1}{1-Q}},
\end{equation}
is a $Q$-Gaussian function, which corresponds to the ordinary exponential in the limit $Q\to 1$:
\begin{equation}
P_0(x,\tau)|_{a\to 0}=P_0(x,\tau_0)\exp [q(\tau-\tau_0)].
\end{equation}

\section{ML Plane-Wave Solutions}\label{SEC:MLPW}

An interesting generalized solution is Mittag-Leffler plave waves. A \textit{chiral solution} refers to the case were the solution depends only on $\xi$ or $\eta$, and not both. Let us consider the simple case \( u_2(\eta) = c \), where \( c \) is a positive constant. Then Eq.~\eqref{Eq.18} reduces to 
  \begin{equation}
  i \omega^\alpha \mathcal{D}^\alpha_\xi u_1 - p k^{2\alpha} \mathcal{D}^{2\alpha}_\xi u_1 - q c^a |u_1|^a u_1 = 0.
  \label{eq.19}
\end{equation}
Note that in the limit $\alpha\to 1$ one retrieves the ordinary plane wave solution of $\exp(i\xi)$. An explicit generalized plane-wave solution for Eq.~\eqref{eq.19} is as follows:
    \begin{equation}
      u_1=e^{iS_{\xi}^{(\alpha)}}=\mathcal{E}_{\alpha}(|\xi|) ,\quad \omega^\alpha=p k^{2\alpha}+q c^a.
    \end{equation}
This solution corresponds to a uniform density $\rho^{(\alpha)}=c^2$
and to a generalized velocity given by (Eq.~\ref{Eq:velocity})
\begin{equation}
v^{(\alpha)}=-2p k^\alpha.
\end{equation}
One tries an independent solution by demanding $u_2=\mathcal{E}_{\alpha}(|\eta|)$. In this case Eq.~\eqref{Eq.18} becomes 
  \begin{equation}
    \begin{split}
        - p &k^{2\alpha} \mathcal{D}^{2\alpha}_\xi u_1 
        + i \left(\omega^\alpha - 2 p r^\alpha k^\alpha \right) \mathcal{D}^\alpha_\xi u_1 \\
        &+ \left(-s^\alpha + p r^{2\alpha} \right) u_1 
        - q \lvert u_1 \rvert^a u_1 = 0.
    \end{split}
    \label{eq21}
\end{equation}

    The explicit solution for Eq.~\eqref{eq21} is
  \begin{equation}
    u_1 = \mathcal{E}_{\alpha}(|\xi|), \quad 
    \omega^\alpha = p (r^\alpha + k^\alpha)^2 - s^\alpha - q.
\end{equation}
where the density is again $\rho^{(\alpha)}=1$, and the classical velocity found to be 
\begin{equation}
v^{(\alpha)}=-2 p(k^\alpha+ r^\alpha).
\end{equation}
\section{A Class of Riccati Soliton Solution}\label{SEC:RiccatiMain}
\begin{figure*}[t]
\centering

\begin{subfigure}[b]{0.4\linewidth}
    \centering
    \includegraphics[width=\linewidth]{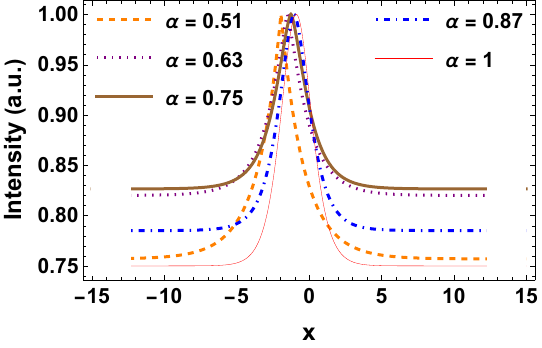}
    
    (a)
\end{subfigure}
\begin{subfigure}[b]{0.4\linewidth}
    \centering
    \includegraphics[width=\linewidth]{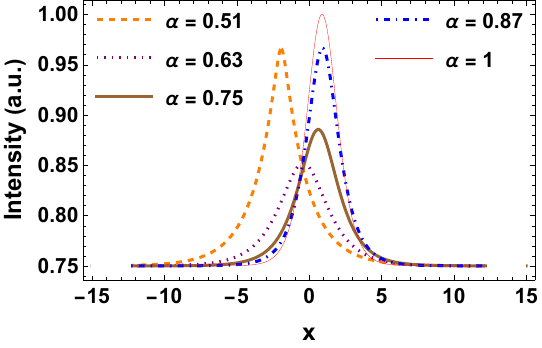}
    
    (b)
\end{subfigure}

\vspace{0.5cm}

\begin{subfigure}[b]{0.4\linewidth}
    \centering
    \includegraphics[width=\linewidth]{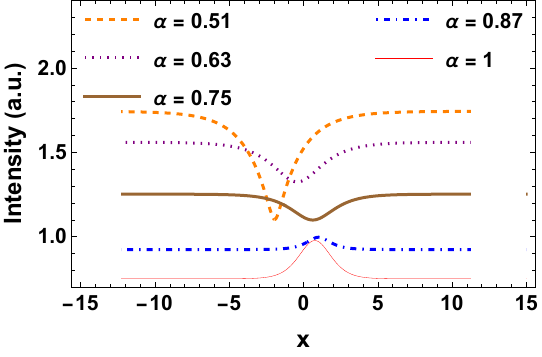}
    
    (c)
\end{subfigure}
\begin{subfigure}[b]{0.4\linewidth}
    \centering
    \includegraphics[width=\linewidth]{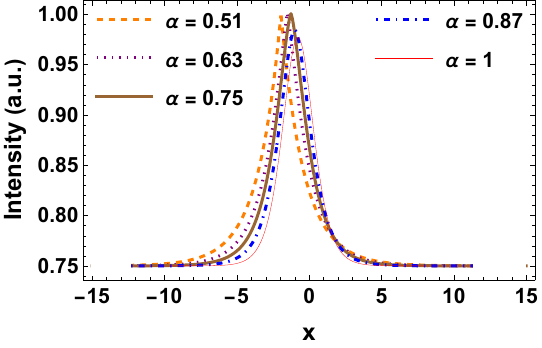}
    
    (d)
\end{subfigure}

\caption{Evolutional behavior of $|u_3|^2$ with $a=2$, $r=0.5$, and $\alpha = 0.51, 0.63, 0.75, 0.87, 1$ at $t = 1$. 
(a) $p=1$, $q=1$; 
(b) $p=-1$, $q=-1$; 
(c) $p=-1$, $q=1$; 
(d) $p=1$, $q=-1$.}
\label{fig:Solitorysol2}
\end{figure*}

\begin{figure*}[t]
\centering

\begin{subfigure}[b]{0.4\linewidth}
    \centering
    \includegraphics[width=\linewidth]{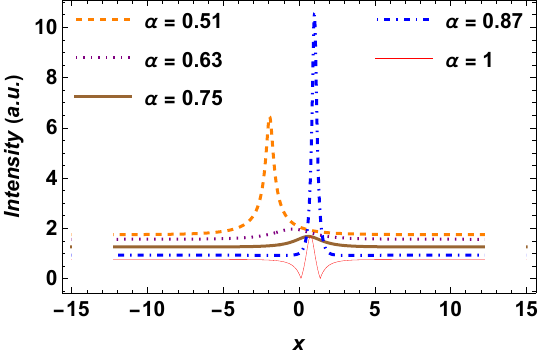}
    
    (a)
\end{subfigure}
\begin{subfigure}[b]{0.4\linewidth}
    \centering
    \includegraphics[width=\linewidth]{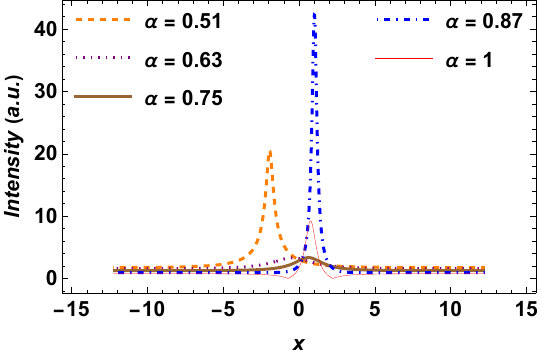}
    
    (b)
\end{subfigure}

\caption{(a) Evolutional behavior of $|u_4|^2$ with $a=2$, $r=0.5$, and $\alpha = 0.51, 0.63, 0.75, 0.87, 1$ at $t = 1$, $p=-1$ and $q=1$; 
~~~(b) Evolutional behavior of $|u_5|^2$ with $a=2$, $r=0.5$, and $\alpha = 0.51, 0.63, 0.75, 0.87, 1$ at $t = 1$, $p=-1$ and $q=1$.}
\label{fig:Solitorysol3}
\end{figure*}

\begin{figure*}[t]
\centering

\begin{subfigure}[b]{0.4\linewidth}
    \centering
    \includegraphics[width=\linewidth]{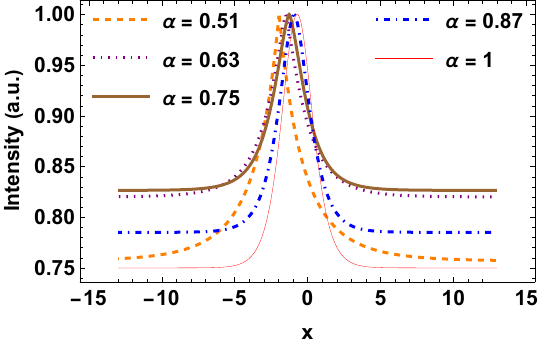}
    
    (a)
\end{subfigure}
\begin{subfigure}[b]{0.4\linewidth}
    \centering
    \includegraphics[width=\linewidth]{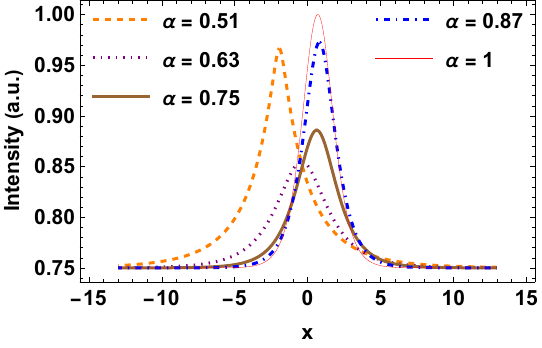}
    
    (b)
\end{subfigure}

\vspace{0.5cm}

\begin{subfigure}[b]{0.4\linewidth}
    \centering
    \includegraphics[width=\linewidth]{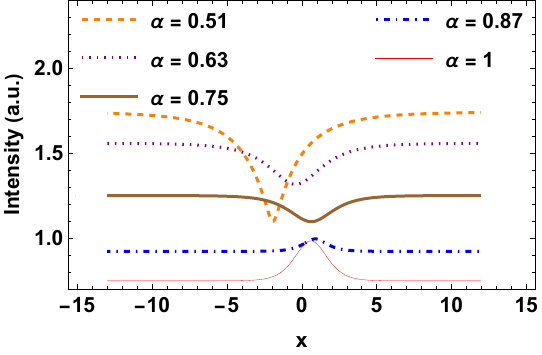}
    
    (c)
\end{subfigure}
\begin{subfigure}[b]{0.4\linewidth}
    \centering
    \includegraphics[width=\linewidth]{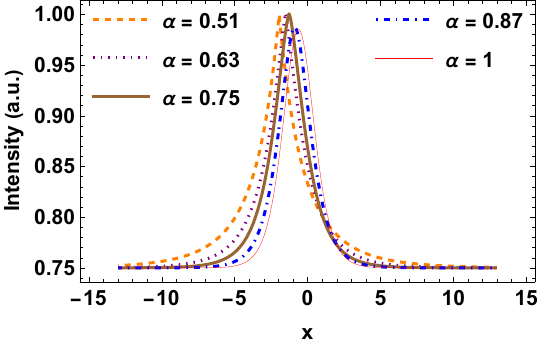}
    
    (d)
\end{subfigure}

\caption{Evolutional behavior of $|u_7|^2$ with $a=2$, $r=0.5$, and $\alpha = 0.51, 0.63, 0.75, 0.87, 1$ at $t = 1$. 
(a) $p=1$, $q=1$; 
(b) $p=-1$, $q=-1$; 
(c) $p=-1$, $q=1$; 
(d) $p=1$, $q=-1$.}
\label{fig:Solitorysol1}
\end{figure*}

In this section, we concentrate on specific class of solutions for FNLSE, referred to as solitons. These distinct solutions characterize localized alterations in the density profile that propagate through the given medium at a uniform velocity, thereby maintaining the inherent shape of the system over time \cite{Pitaevskii2003}.
We consider the solitons obtained for the Eq.~\eqref{eq21}, i.e. $u_2=\mathcal{E}_\alpha(|\eta|)$. We use the fractional Riccati method, and therefore name the corresponding solutions as the Ricatti solitons. We consider the following transformation
\begin{equation}
     u_1(\xi)= \phi^{\frac{1}{a}}(\xi). \label{eq23}
\end{equation}
Substituting Eq.~\eqref{eq23} into Eq.~\eqref{eq21} and multiplying by \( a^2 \phi^{\frac{2a-1}{a}} \), we obtain
\begin{equation}
\begin{split}
    &~~-p k^{2\alpha} \left[ (1-a)(\mathcal{D}^\alpha_\xi \phi)^2 - a \phi \mathcal{D}^{2\alpha}_\xi \phi \right] 
    - q~a^2 \phi^3\\
    &+ i \left( w^\alpha - 2 p r^\alpha k^\alpha \right) a \phi \mathcal{D}^\alpha_\xi \phi 
    + (p r^{2\alpha} - s^\alpha) a^2 \phi^2 = 0.
\end{split}
\label{eq17}
\end{equation}
As described in SEC.~\ref{SEC:RicattiMain}, we balance terms to find a best choice for $N$. Follwoing this procedure, we balance the term \( \phi^3 \) with \( \phi \mathcal{D}^{2\alpha}_x \phi \) in Eq.~\eqref{eq17}, resulting to \( N = 2 \), for the details see SEC.~\ref{SEC:RicattiMain} and appendix~\ref{SEC:Ricatti}. Therefore, $\phi(\zeta)$ is expanded as
\begin{equation}
\phi(\xi) = a_0 + a_1 F(\xi) + a_2 F(\xi)^2, \label{sol}
\end{equation}
%where
%\begin{equation}
%\begin{split}
%a_0 &= 1, \quad a_1 = 0, \quad a_2 = \frac{2 (p r^{2\alpha} - s^\alpha) B q}{(2+a)A}, \\
%k^{2\alpha} &= \frac{-a^2 q}{2(2+a) p AB}, \quad s^\alpha - p r^{2\alpha} = \frac{-2q}{(2+a)}, \\
%w^\alpha &= 2 p r^\alpha k^\alpha.
%\end{split}
%\end{equation}
where the coefficients \( a_0, a_1, a_2 \) are unknown multipliers to be found using the Riccati method, along with the unknown dispersion pairs $(k,\omega)$ and $(r,s)$ that should be obtained by solving the corresponding algebraic equations. These coefficients are found to be $a_0 = 1$, $a_1=0$, and
\begin{equation}
a_2\equiv a_2(A,B)=\frac{2Bq (p r^{2\alpha} - s^\alpha)}{A(2 + a)},
\end{equation}
where $A$ and $B$ are the coefficients introduced in Eq.~\ref{Eq:Fequation}, which tune the soliton solutions, see Eq.~\ref{???} in Appendix~\ref{SEC:Ricatti} for the different choices.
The final general solution is then found to be
\begin{equation}
u = \left[a_0 + a_2(A,B) F(\xi)^2 \right]^{\frac{1}{a}} 
\mathcal{E}_{\alpha}(|\eta|),
\label{eq:2.14}   
\end{equation}
One finds the following dispersion relations for $(k,\omega)$:
\begin{equation}
 w^\alpha = 2 p r^\alpha k^\alpha,  
\end{equation}
and also for $(r,s)$:
\begin{equation}
  s^\alpha  = \frac{-2q}{(2 + a)}+ p r^{2\alpha}.  
\end{equation}
The amount of $k$ depends on the choice of $A$ and $B$ and the Hamiltonian parameters $(p,q,a)$ as follows:
\begin{equation}
k^{2\alpha} = \frac{-a^2 q}{2(2 + a) p AB}.\label{dispersionk}
\end{equation}
Based on the Riccati method, which admits different independent solution with different choices of \( A \) and \( B \), we find $9$ solitoray solutions. These solutions end up with different forms of \( F \), called fractional trigonometric and hyperbolic functions:
\begin{equation}
\begin{split}
\tanh(\xi,\alpha) &=\frac{e_\alpha(\xi)^2-1}{e_\alpha(\xi)^2+1},\ 
\tan(\xi,\alpha) =\frac{\mathcal{E}_\alpha(\xi)^2-1}{\mathcal{E}_\alpha(\xi)^2+1}.
\end{split}
\end{equation}
see Appendix \ref{SEC:Ricatti} for the details, and the definition of the other functions. The solutions are classified as follows
\begin{equation}
u^{(i)}(\xi,\eta) = \left[ 1 + m_ia_2(1,1) f_i(|\xi|, \alpha) \right]^{\frac{1}{a}} \mathcal{E}_{\alpha}(|\eta|).
\end{equation}
where $i=1,...,9$ is the number of solutions, and
\begin{equation}
k_i^{2\alpha} = C_i \frac{a^2 q}{2(2 + a)p},
\end{equation}
and the coefficients are defined as 
\begin{equation}
\begin{split}
&-C_{1,2} =C_{3,4}=\frac{1}{4}C_5=\frac{-1}{4}C_6=4 C_7=-4 C_{8,9}= 1,\\
&m_{1,2,6}=- m_{3,4,5}=\frac{-1}{4}m_7=\frac{1}{4}m_{8,9}=1.  
\end{split}
\end{equation}
In these relations, the fractional trigonometric and hyperbolic functions are defined as
\begin{equation}
\begin{aligned}
& f_{1,2,3,4} = \{\tan (\xi,\alpha), \cot(\xi,\alpha),\tanh(\xi,\alpha), \coth(\xi,\alpha)\}, \\
& f_5 = f_4 \pm \text{csch} (\xi,\alpha), \quad f_6 = \csc(\xi,\alpha) + f_2, \\
& f_7 = \frac{f_3}{1 + f_3^2},\quad f_8 = \frac{f_1}{1 - f_1^2}, \quad
f_9 = \frac{f_2}{1 - f_2^2}.
\end{aligned}
\end{equation}
Not all the solutions obtained via the Riccati method are physically meaningful, as some exhibit singular behavior. The explicit forms of the non-singular solitonic solutions are presented in Table~\ref{tab:solutions} and illustrated in Figs.\ref{fig:Solitorysol2}, \ref{fig:Solitorysol3}, and~\ref{fig:Solitorysol1}. In these representations, the parameters $p$ and $q$ are set to $\pm 1$. While they can, in principle, be any positive or negative values, the equation can be rescaled such that $p$ and $q$ are normalized to $\pm 1$.\\

An interesting feature of the solutions we found are the \textit{focusing-defocusing transition} in terms of the fractionality parameter $\alpha$ for $p=-1$ and $q=1$. More precisely, as shown in Figs.~\ref{fig:Solitorysol2} and~\ref{fig:Solitorysol1}, the solutions generally exhibit bright soliton profiles for all chosen values of \( p \) and \( q \), except in the case of \( p = -1 \) and \( q = 1 \), where a dark soliton gradually transforms into a bright soliton as the parameter \( \alpha \) increases. A dark soliton is a localized drop in intensity (a \textit{dip}) in a continuous wave background, while a bright soliton is a localized pulse of wave amplitude that decays to zero or a constant at spatial infinity. Physically, this indicates that the balance between nonlinearity and dispersion shifts with increasing \( \alpha \), altering the nature of the soliton. In contrast, Figure~\ref{fig:Solitorysol3} demonstrates that for \( p = -1 \) and \( q = -1 \), the solutions consistently remain in the form of bright solitons for all values of \( \alpha \).\\
\begin{table*}[t]
\centering
\begin{tabular}{|c|c|c|c|c|c|}
\hline
$(n,p,q)$ & \textbf{stability} & $u^{(n)}(\xi,\eta)^a\times \mathcal{E}(|\eta|)^{-a}$ & \textbf{$s^\alpha$}&$\omega^\alpha$ & $k^{2\alpha}$ \\
\hline
(3,1,1) (Fig.~\ref{fig:Solitorysol2}a)  & \checkmark (Fig.~\ref{fig:stability1}a) & $ 1 +\frac{-2(r^{2\alpha}-s^\alpha)}{2+a} \tanh^2(|\xi|, \alpha) $ & $\frac{-2}{2 + a} + r^{2\alpha}$ & $2r^\alpha k^\alpha$ & \multirow{2}{*}{$ \frac{a^2}{2(2+a)}$} \\
\cline{1-5}
(3,-1,-1) (Fig.~\ref{fig:Solitorysol2}b) & \checkmark (Fig.~\ref{fig:stability1}b) & $ 1 +\frac{2(-r^{2\alpha}-s^\alpha)}{2+a} \tanh^2(|\xi|, \alpha) $ & $\frac{2}{2 + a} - r^{2\alpha}$ & $-2r^\alpha k^\alpha$ & \\
\hline
(3,-1,1) (Fig.~\ref{fig:Solitorysol2}c) & \checkmark (Fig.~\ref{fig:stability1}c) & $ 1 +\frac{2(r^{2\alpha}-s^\alpha)}{2+a} \tanh^2(|\xi|, \alpha) $ & $\frac{2}{2 + a} + r^{2\alpha}$ & $2r^\alpha k^\alpha$ & \multirow{3}{*}{$ \frac{-a^2}{2(2+a)}$} \\
\cline{1-5}
(3,1,-1) (Fig.~\ref{fig:Solitorysol2}d) & \checkmark (Fig.~\ref{fig:stability1}b) & $1 +\frac{-2(-r^{2\alpha}-s^\alpha)}{2+a} \tanh^2(|\xi|, \alpha) $ & $\frac{-2}{2 + a} - r^{2\alpha}$ & $-2r^\alpha k^\alpha$ & \\
\cline{1-5}
(4,-1,1) (Fig.~\ref{fig:Solitorysol3}a) & \checkmark (Fig.~\ref{fig:stability3}a) & $1 +\frac{-2(-r^{2\alpha}-s^\alpha)}{2+a} \coth^2(|\xi|, \alpha) $ & $\frac{-2}{2 + a} - r^{2\alpha}$ & $-2r^\alpha k^\alpha$ & \\
\hline
(5,-1,1) (Fig.~\ref{fig:Solitorysol3}b) & \checkmark (Fig.~\ref{fig:stability3}b) & $1 +\frac{-2(-r^{2\alpha}-s^\alpha)}{2+a} \left(\coth(|\xi|, \alpha)+ \operatorname{csch}(|\xi|, \alpha)\right)^2$ & $\frac{-2}{2 + a} - r^{2\alpha}$ & $-2r^\alpha k^\alpha$ & $\frac{-2a^2}{2+a}$ \\
\hline
(7,1,1) (Fig.~\ref{fig:Solitorysol1}a) & \checkmark (Fig.~\ref{fig:stability4}a) & $1 +\frac{-8(r^{2\alpha}-s^\alpha)}{2+a} \left(\frac{\tanh(|\xi|, \alpha)}{1+\tanh^2(|\xi|, \alpha)}\right)^2$ & $\frac{-2}{2 + a} + r^{2\alpha}$ & $2r^\alpha k^\alpha$ & \multirow{2}{*}{$ \frac{a^2}{8(2+a)}$} \\
\cline{1-5}
(7,-1,-1) (Fig.~\ref{fig:Solitorysol1}b) & \checkmark (Fig.~\ref{fig:stability4}b) & $1 +\frac{8(-r^{2\alpha}-s^\alpha)}{2+a} \left(\frac{\tanh(|\xi|, \alpha)}{1+\tanh^2(|\xi|, \alpha)}\right)^2$ & $\frac{2}{2 + a} - r^{2\alpha}$ & $-2r^\alpha k^\alpha$ & \\
\hline
(7,-1,1) (Fig.~\ref{fig:Solitorysol1}c) & \checkmark (Fig.~\ref{fig:stability4}c) & $1 +\frac{-8(-r^{2\alpha}-s^\alpha)}{2+a} \left(\frac{\tanh(|\xi|, \alpha)}{1+\tanh^2(|\xi|, \alpha)}\right)^2$ & $\frac{-2}{2 + a} - r^{2\alpha}$ & $-2r^\alpha k^\alpha$ & \multirow{2}{*}{$ \frac{-a^2}{8(2+a)}$} \\
\cline{1-5}
(7,1,-1) (Fig.~\ref{fig:Solitorysol1}d) & \checkmark (Fig.~\ref{fig:stability4}d) & $1 +\frac{8(r^{2\alpha}-s^\alpha)}{2+a} \left(\frac{\tanh(|\xi|, \alpha)}{1+\tanh^2(|\xi|, \alpha)}\right)^2$ & $\frac{2}{2 + a} + r^{2\alpha}$ & $2r^\alpha k^\alpha$ & \\
\hline
\end{tabular}
\caption{Non-singular solutions for different values of $n$ $p$ and $q$ and corresponding dispersion relations. The stability test is based on Eq.~\ref{Eq:stability}. The symbol (\checkmark) shows that the solution is stable in terms of the relation found for $k$.}
\label{tab:solutions}
\end{table*}
%==========================
%========================
\begin{figure*}[t]
\centering

\begin{subfigure}[b]{0.4\linewidth}
    \centering
    \includegraphics[width=\linewidth]{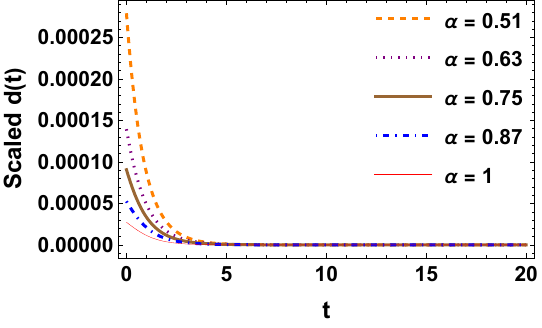}
    
    (a)
\end{subfigure}
\begin{subfigure}[b]{0.4\linewidth}
    \centering
    \includegraphics[width=\linewidth]{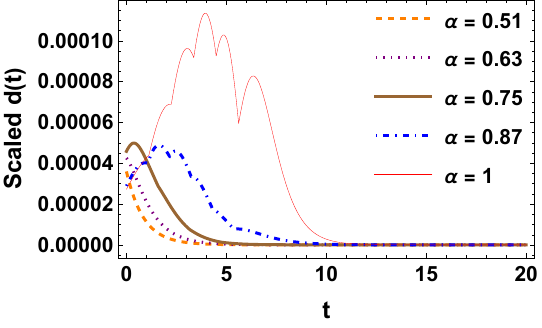}
    
    (b)
\end{subfigure}

\vspace{0.5cm}

\begin{subfigure}[b]{0.4\linewidth}
    \centering
    \includegraphics[width=\linewidth]{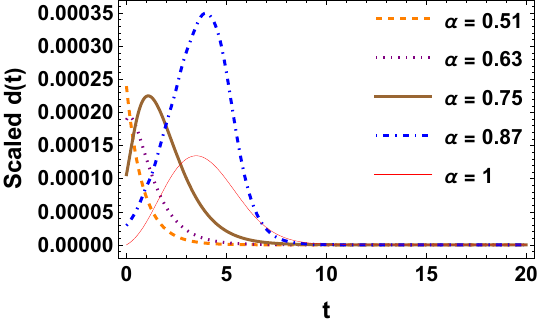}
    
    (c)
\end{subfigure}
\begin{subfigure}[b]{0.4\linewidth}
    \centering
    \includegraphics[width=\linewidth]{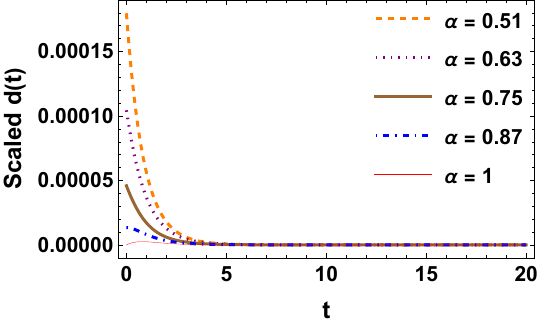}
    
    (d)
\end{subfigure}

\caption{Stability test  of $u_3$ with $r=0.5$, and $\alpha = 0.51,0.63,0.75,0.87,1$. 
(a) $p=1$, $q=1$; 
(b) $p=-1$, $q=-1$; 
(c) $p=-1$, $q=1$; 
(d) $p=1$, $q=-1$.}
\label{fig:stability1}
\end{figure*}

\begin{figure*}[t]
\centering

\begin{subfigure}[b]{0.4\linewidth}
    \centering
    \includegraphics[width=\linewidth]{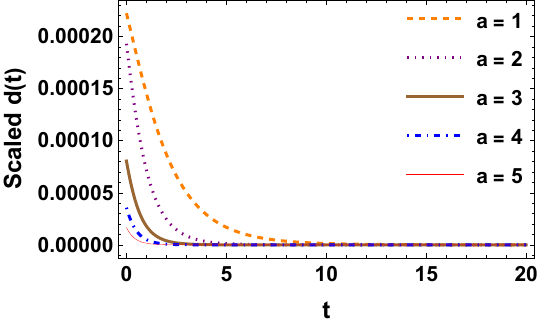}
    
    (a)
\end{subfigure}
\begin{subfigure}[b]{0.4\linewidth}
    \centering
    \includegraphics[width=\linewidth]{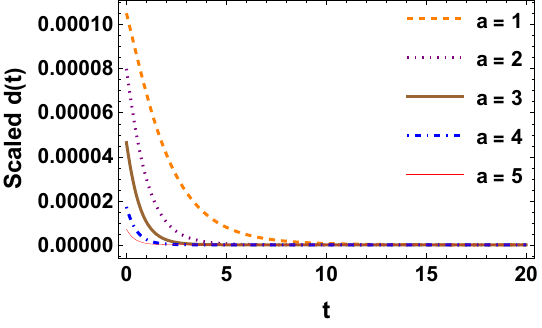}
    
    (b)
\end{subfigure}

\vspace{0.5cm}

\begin{subfigure}[b]{0.4\linewidth}
    \centering
    \includegraphics[width=\linewidth]{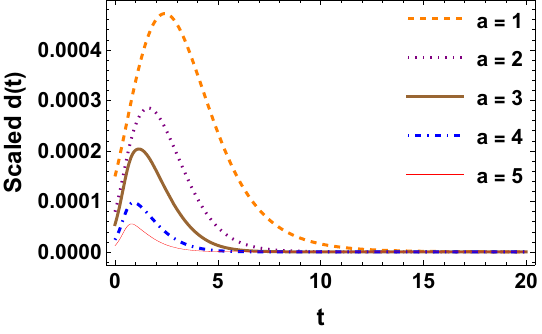}
    
    (c)
\end{subfigure}
\begin{subfigure}[b]{0.4\linewidth}
    \centering
    \includegraphics[width=\linewidth]{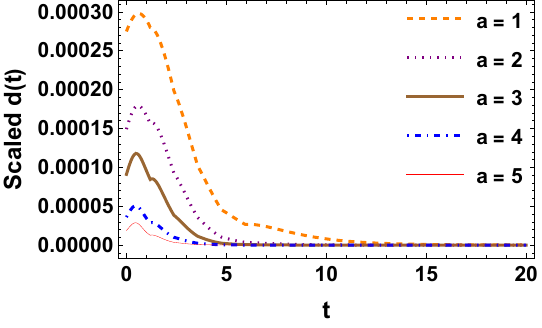}
    
    (d)
\end{subfigure}

\caption{Stability test  of $u_3$ with $r=0.5$, and $\alpha = 0.8$. 
(a) $p=1$, $q=1$; 
(b) $p=1$, $q=-1$; 
(c) $p=-1$, $q=1$; 
(d) $p=-1$, $q=-1$.}
\label{fig:stability2}
\end{figure*}

From Table~\ref{tab:solutions}, we observe that the soliton solutions are found for especial amounts of $k$, which arises the question of whether changing slightly the amount of $k$ destroys the soliton solution or not, i.e. the \textit{stability of the solution}. We have analyzed this issue in Appendix~\ref{Appendix:stability} in which we address thoroughly the stability of all the non-singular solutions presented in table~\ref{tab:solutions} and for various rates of $\alpha$ and $a$. The stability is expressed in terms of distance between the solitonic solution, and the \textit{neighboring} solutions. To this end, we have defined a distance between the soliton solution and the perturbed solution as follows (see the equations leading to Eq.~\ref{Eq:stability}):
\begin{equation}
d_k^{(n)}(t)\equiv\int_{-\infty}^\infty\text{d}x\left|u^{(n)}_{k+\delta k}(x,t)-u^{(n)}_{k}(x,t)\right|,
\label{Eq:stabilityMain}
\end{equation}
where $k$ is the exact amount proposed in Eq.~\ref{dispersionk} and table~\ref{tab:solutions}. In this equation we calculate the absolute value of the differences, integrated over the whole space. The results are presented in Figs.~\ref{fig:stability1} and~\ref{fig:stability2} for various amounts of $p$ and $q$ ($\delta k=0.001$), from which we observe that all the solutions we reported in table~\ref{tab:solutions} are stable in terms of $k$. This demonstrates that the solutions with neighboring $k$ values return to the soliton solution in large time limit, i.e. if we give enough time to the system.  
%====================================
\section{Perturbative Adomian Solutions}\label{SEC:Adomian}
In this section, we present several perturbative solutions obtained using the Adomian decomposition method. These solutions naturally depend on the initial form of the functions involved. The following analysis illustrates examples of how the Adomian decomposition method can be applied within a general framework.
\subsection{Chiral Solution}\label{SEC:chiral}
We consider a chiral solution Eq. \eqref{eq.19}, and use the Adomian decomposition framework. More precisely, we consider the case where $u_2(\eta)=c$ is a constant. Applying the inverse operator $\mathcal{J}^\alpha_0$ (Eq.~\ref{Eq:RLIntegral}) to its rearranged form gives
\begin{equation} u_1(\xi) = u_1(0) - i \mathcal{J}^\alpha_0 \left( p \frac{k^{2\alpha} }{w^\alpha} \mathcal{D}^{2\alpha}_\xi u_1 + q \frac{c^a}{w^\alpha} |u_1|^a u_1 \right).\label{50} \end{equation}
Then the solution is expanded as
\begin{equation}
u_1(\xi)=\sum_{j=0}^{\infty}u_1^{(j)}(\xi),\label{uexpansion}
\end{equation}
where $u_1^{(n)}(\xi)$ is the Adomian perturbative term in the $n$th order. The Eq.~\eqref{50} leads to the following recursive relation:
\begin{equation}
\begin{split}
   &u_1^0(\xi)=u_1(0) =z,\\
   &u_1^{(n+1)}(\xi)=- i \mathcal{J}^\alpha_0 \left( p \frac{k^{2\alpha} }{w^\alpha} \mathcal{D}^{2\alpha}_\xi u_1^{(n)} + q \frac{c^a}{w^\alpha} A_n\right).
\end{split}
\end{equation}
where $z$ is a positive constant, and $A_n$ defined in Eq.~\ref{App:C1} is the Adomian non-linear term:
\begin{equation}
A_n = \frac{1}{n!} \left. \frac{\text{d}^n}{\text{d}\lambda^n} \mathcal{N}\left(\sum_{j=0}^n u_j \lambda^j\right) \right|_{\lambda=0}.
\label{Eq:Adomian}
\end{equation}
Note that $\mathcal{N}$ represents the non-linear term in the equation, which is $\mathcal{N}(y)=|y|^ay$ in our case. The Adomian terms are then found to be
\begin{equation}
\begin{split}
   &u_1^{(0)} = z, \\
   & u_1^{(1)} = -i z \frac{q c^a z^{a}}{w^\alpha \Gamma(\alpha+1)} \xi^\alpha, \\
   &u_1^{(2)} =-z \frac{q^2 c^{2a} z^{2a}}{w^{2\alpha} \Gamma(2\alpha+1)} \xi^{2\alpha}, \\&
  u_1^{(3)} = i z \left( \frac{p q^2 k^{2\alpha} c^{2a} z^{2a}}{\omega^{3\alpha} \Gamma(\alpha+1)} \xi^\alpha + \frac{\kappa q^3 c^{3a} z^{3a}}{\omega^{3\alpha} \Gamma(3\alpha+1)}  \xi^{3\alpha} \right) 
\\&\dots
\end{split}
\end{equation}
where $\kappa = \left( a+1
    - \frac{a \Gamma(2\alpha+1)}{2 \Gamma(\alpha+1)^2 } \right)$. Based on this decomposition, we propose the following series expansion
\begin{equation}
\begin{split}
u_1(\xi)=&\sum_{j\ge 0}\frac{z(-i)^j l_j\xi^{j\alpha}}{\Gamma(j\alpha+1)}
\end{split}
\label{Eq:Adomian1}
\end{equation}
where $l_j$ is a $j$th order real-valued coefficient which depends generally on $\alpha$, $c$, $z$, $a$, $p$, $q$ and $\omega$.
\subsection{A Plane-Wave Perturbative Adomian Solution}\label{SEC:PWAdomian}
In this subsection, we consider another possibility according to SEC.~\ref{SEC:RiccatiMain}, i.e. the case where $u_2(\eta)= \mathcal{E}_\alpha(|\eta|)$ in Eq.\eqref{Eq.18}, which leads to Eq.~\eqref{eq21}, using the Adomian decomposition method. After isolating the term \( \mathcal{D}_\xi^\alpha u \) on the left-hand side of Eq.~\eqref{eq21} and moving the remaining terms to the right-hand side, we apply the inverse operator \( \mathcal{J}^\alpha_0 \) to both sides as follows:
 \begin{equation}
\begin{split}
    u_1(\xi) = u&_1(0) - \frac{i}{w^\alpha - 2 p r^\alpha k^\alpha} \mathcal{J}^\alpha_0 \Big[ 
    p k^{2\alpha} \mathcal{D}^{2\alpha}_\xi u_1(\xi) \\&\quad+ (s^\alpha - p r^{2\alpha}) u_1(\xi) 
     + q |u_1(\xi)|^a u_1(\xi) \Big].    \label{70}
\end{split}
\end{equation}
Here, we assume that \( w^\alpha \neq 2pr^\alpha k^\alpha \); otherwise, the corresponding term can be considered trivial and thus neglected. Subsequently, we compute the Adomian decomposition by applying the inverse operator \( \mathcal{J}^{2\alpha} \). In the first case, the solution is expanded as Eq.\eqref{uexpansion}, resulting in the following recursive relation:
\begin{equation}
\begin{aligned}
   &u_1^{(0)}(\xi)=u_1(0) =z \\
   &u_1^{(n+1)} = \frac{-i}{\omega^\alpha - 2 p r^\alpha k^\alpha} \mathcal{J}^\alpha_\xi \Big[ 
    p k^{2\alpha} \mathcal{D}^{2\alpha}_\xi u_1^{(n)}(\xi)  \\
    &\quad~~~~~~~~~~~~~~~~~ + (s^\alpha - p r^{2\alpha}) u_1^{(n)}(\xi) + q A_n \Big].
\end{aligned}
\end{equation}
 as in the previous subsection, \( z \) is a positive constant and \( A_n \) denotes the Adomian component, as defined in Eq.~\eqref{Eq:Adomian}. The first few terms in the series are found to be
\begin{equation}
\begin{split}
 &u_1^{(0)} = z, \\
 & u_1^{(1)} = -i z \left( \frac{s^\alpha - p r^{2\alpha} +q z^{a}}{\omega^\alpha - 2 p r^\alpha k^\alpha } \right)\frac{\xi^\alpha}{\Gamma(\alpha+1)}, \\
&u_1^{(2)} = -z \left( \frac{s^\alpha - p r^{2\alpha} + q z^a}{\omega^\alpha - 2 p r^\alpha k^\alpha} \right)^2 
\frac{\xi^{2\alpha}}{\Gamma(2\alpha+1)}
,\\& u_1^{(3)} = iz \Bigg( 
\frac{p k^{2\alpha} \left( s^\alpha - p r^{2\alpha} + q z^a \right)^2}{\left( \omega^\alpha - 2 p r^\alpha k^\alpha \right)^3} \frac{\xi^{\alpha}}{\Gamma(\alpha+1)}
\\
& + \Lambda \left( \frac{ s^\alpha - p r^{2\alpha} + q z^a }{ \omega^\alpha - 2 p r^\alpha k^\alpha } \right)^2 \frac{\xi^{3\alpha}}{\Gamma(3\alpha+1)}
\Bigg), 
\\&\dots  
\end{split}
\label{Eq:Adomian2}
\end{equation}
where $\Lambda= s^\alpha- pr^{2\alpha}+a+1+\frac{a \Gamma(2\alpha+1)}{2 \Gamma(\alpha+1)^2}$.
The solution for \( u_1 \) is obtained by summing the components above and the subsequent terms. While the structure of this summation resembles that in Eq.~\ref{Eq:Adomian1}, the coefficient \( l_j \) used here differs from the one in \ref{Eq:Adomian1}. In the present context, \( l_j \) is a real-valued coefficient at order $j$ determined by \( \alpha \), \( z \), \( a \), \( p \), \( q \), and \( \omega \).
\section{Self-Similar Solutions}\label{SEC:SelfSimilar} 
In this section, we investigate self-similar solutions of the Eq.~\eqref{Eq:NLSE} using scaling laws. We begin by applying the following scaling transformations to  Eq.~\eqref{Eq:NLSE}:
\begin{equation}
\begin{split}
& t \rightarrow \lambda t, \quad x \rightarrow \lambda^{\beta'} x, \quad u \rightarrow \lambda^{-\eta'} u,\\
&\mathcal{D}^\alpha_t \rightarrow \lambda^{-\alpha} \mathcal{D}^\alpha_t, \quad \mathcal{D}^{\beta}_x \rightarrow \lambda^{-\beta\beta'} \mathcal{D}^{\beta}_x,\label{transformation}
 \end{split}   
\end{equation}
where $(\beta',\eta')$ are the scaling exponents and $(\beta,\alpha)$ are the order of fractional derivative. Although $\beta=2\alpha$, we keep it general until the end of calculations. By applying the scaling transformation, one gets
\begin{equation}
    \lambda^{-\eta'-\alpha} \left( i \mathcal{D}^{\alpha}_t u \right) -p \lambda^{-\eta'-\beta \beta'} \mathcal{D}^{\beta}_x u -q \lambda^{-\eta' a -\eta'} |u|^a u = 0.\label{scaled}
\end{equation}
 To ensure a scale-invariant solution, the powers of 
$\lambda$ in each term must match. This leads to the following relations: 
\begin{equation}
    \beta' = \frac{\alpha}{\beta}, \quad \eta' = \frac{\alpha}{a}.\label{scalingexponents}
\end{equation}
We consider the following trial scale-invariant solution:
\begin{equation}
u(x, t) = t^{-\eta'} \Phi\left( \frac{x}{t^{\beta'}} \right), 
\label{Eq:scalingMain}
\end{equation}
where \( \Phi(\chi) \) is a function of the scaling variable \( \chi = \frac{x}{t^{\beta'}} \), and the scaling exponents \( \eta' \) and \( \beta' \) are defined in Eq.~\eqref{scalingexponents}.
We then substitute Eq.~\eqref{Eq:scalingMain} into Eq.~\eqref{Eq:NLSE}, and use key properties of the Modified RL derivative, specifically Eqs. \eqref{eq.A2} ,\eqref{eq.A4} and \eqref{eq:A6}, and also using the fact that $\eta'(a+1) = \frac{\alpha}{a}(a+1) = \alpha + \eta'$ and $\eta' + \beta \beta' = \eta' + \alpha$ which results in:
\begin{equation}
    i \mathcal{G} \Phi(\chi) 
    + i (-\beta' \chi)^\alpha \mathcal{D}^\alpha_\chi \Phi(\chi) 
    - p \mathcal{D}^{\beta}_\chi \Phi(\chi) 
    - q |\Phi(\chi)|^a \Phi(\chi) = 0.
    \label{eq97}    
\end{equation}
where \( \mathcal{G} = \frac{\Gamma(1-\eta')}{\Gamma(1-\eta'-\alpha)} \). Solving the Eq.\eqref{eq97} can help us to  find self-similar solution to Eq.\eqref{Eq:NLSE}, which provides extensive insights into the system's behavior.
In the following sections, we investigate solutions to Eq.~\eqref{eq97} based on the Mittag-Leffler exponential function and the Adomian decomposition method.
\subsection{A Self-Similar Pure-Phase Solution}
In this subsection, we find a solution based on the Mittag-Leffler exponential to Eq.~\eqref{eq97}. One explicit solution is found to be:
\begin{equation}
  \Phi(\chi)=\mu ~~\mathcal{E}_{\alpha}\left(\left| \frac{w \chi^2}{2}\right|\right), \end{equation}
  where $\mu^a =  \frac{i \left( \mathcal{G}
    - p w^{\alpha} \Gamma(\alpha+1) \right)}{q}$, and $w^\alpha=\frac{(-\beta')^\alpha}{p}$.  This formulation provides a self-similar description of the FNLSE solution in terms of the Mittag-Leffler function \( \mathcal{E}_\alpha \), highlighting the role of fractional-order dynamics in the system.  Note that, for the stochastic dual system, this solution corresponds to the generalized Gaussian distribution Eq.~\ref{Eq:GaussianMain} using the generalized Wick's rotation Eq.~\ref{Eq:WickRotation}.
\subsection{A Perturbative Adomian Solution}\label{SEC:ScalingAdomian}
In this subsection, we compute perturbatively the expansion components of the function $\Phi(\chi)$ in Eq.~\eqref{eq97} using the Adomian decomposition method. We first arrange the Eq.~\eqref{eq97} by keeping $\mathcal{D}^\alpha_\chi \Phi(\chi)$ in the left-hand side and remaining terms in the right side, and then apply the inverse operator \( \mathcal{J}^\alpha_0 \) to both sides. The following equation is obtained:
\begin{equation}
    \begin{aligned}
        \Phi(\chi) &= \Phi(0) 
        - i \mathcal{J}^\alpha_0 \Big[ (-\beta' \chi)^{-\alpha} \Big( 
        -i \mathcal{G} \Phi(\chi) \\
        &~~~~~~~\quad + p \mathcal{D}^\beta_\chi \Phi(\chi) 
        + q |\Phi(\chi)|^a \Phi(\chi) \Big) \Big].   
    \end{aligned}
\end{equation}
According to the main strategy, one should expand \( \Phi(\chi) \) as \( \sum_{j=0}^{\infty} \Phi_j(\chi) \) and then express \( |\Phi|^a \Phi \) in terms of Adomian polynomials. We note that for large enough times $t\to\infty$, small $\chi$ values are of interest. Therefore, we are interested in small $\chi$ values in this limit, so that the \textit{initial value} of $\Phi$ should be considered in the leading order. We consider the case $\Phi_0(\chi)=z_0+z_1\chi^\alpha$, where $z_0$ and $z_1$ are two real numbers. This gives the following recursive relation is obtained:
\begin{equation}
    \begin{aligned}
        &\Phi_0(\chi) = z_0-z_1\chi^\alpha,  \\
        &\Phi_{n+1}(\chi) = - i \mathcal{J}^\alpha_0 \Big[ (-\beta' \chi)^{-\alpha} 
        \Big( -i \mathcal{G} \Phi_n(\chi) 
        \\&~~~~~~~~~~~~~~~~~~~~~~~+ p \mathcal{D}^\beta_\chi \Phi_n(\chi) 
        + q A_n \Big) \Big].   
    \end{aligned}
\end{equation}

In the following we keep all the calculations up to $O(\chi^\alpha)$, and use the fact that $\left| 1- \frac{z_1}{z_0} \chi^\alpha \right|^a \approx 1- \frac{z_1 a}{z_0} \chi^\alpha$.
We also consider the case of our interest \( \beta = 2\alpha \), which consequently gives \( \beta'= \frac{1}{2} \). The components are then found to be:\\
\begin{widetext}
\begin{equation}
    \begin{aligned}
        \Phi_0(\chi) &= z_0- z_1 \chi^\alpha, \ \Phi_1(\chi) = 2^{\alpha} e^{-i \alpha \pi} 
\left[
\left( -\mathcal{G} z_0 - i q z_0^{a+1} \right) \Gamma(1 - \alpha)
+ \left( \frac{\mathcal{G} z_1 - 2i q z_0^a z_1}{\Gamma(\alpha + 1)} \right) \chi^\alpha
\right],\\
 2^{-2\alpha}\Phi_2(\chi) &=\left[  e^{-2i\alpha \pi}\left(-\mathcal{G}-i q(\frac{a}{2}+1) z_0^a \right) \left(-\mathcal{G}  z_0 
 -i q z_0^{a+1}\right)-i q \frac{a}{2}  \left( -\mathcal{G}  z_0^{a+1} +i q z_0^{2a+1} \right) \right] \Gamma(1-\alpha)^2\\ &\quad+ \left[ e^{-2i\alpha \pi}\left( -\mathcal{G}-i q (\frac{a}{2}+1) z_0^a \right) \left(\frac{\mathcal{G} z_1 +2i q z_0^a  z_1 }{\Gamma(\alpha+1)^2}\right)- iq  \frac{a}{2} \left( \frac{\mathcal{G} z_1 z_0^a -2iq z_0^{2a}z_1}{\Gamma(\alpha+1)^2}\right)\right]\chi^\alpha
\\ &\quad -i q z_0^{a-1} z_1 \frac{\Gamma(1-\alpha)}{\Gamma(1+\alpha)} \left[   e^{-2 i \alpha \pi}  (\frac{a}{2}+1) a \left( \mathcal{G}  z_0 +i q z_0^{a+1} \right)+ \frac{a}{2}\left( \mathcal{G}  z_0 -i q z_0^{a+1} \right) \right] \chi^\alpha,
 \\&\dots
    \end{aligned}
    \label{Eq.scalingAdomian}
\end{equation}  
\end{widetext}
where \( \mathcal{G} = \frac{\Gamma(1-\eta')}{\Gamma(1-\eta'-\alpha)} \), and $z$ is a positive initial value constant, and the calculations are kept up to order $O(\chi^\alpha)$. This expansion is useful as it captures the system's asymptotic behavior in the limit $t\to \infty$, where $\chi\to 0$, making a few leading terms sufficient.

\section{Concluding Remarks}
In this work, we have conducted a comprehensive analytical study of the space-time fractional nonlinear Schrödinger equation (FNLSE) incorporating the modified Riemann–Liouville derivative as formulated by Jumari. The FNLSE is characterized by two parameter: the fractional parameter ($\alpha$, which captures the memory effects) and the non-linearity parameter ($a$). The FNLSE is a natural generalization of the conventional NLSE that captures essential features of memory effects, and nonlocal dynamics, making it an effective tool for modeling complex wave propagation in nonlinear media with fractional spatial and temporal correlations. To address the mathematical and physical richness of the FNLSE, we employed three distinct but complementary techniques: the fractional Riccati method, the Adomian decomposition method, and the scaling method. These methods enabled us to construct exact or approximate solutions of various qualitative types—most notably bright and dark solitons, chiral plane waves, and self-similar waveforms. Each method offers unique insights into the underlying structure of the FNLSE and facilitates classification of solution families in different functional and dynamical regimes.

We began our analysis by recasting the FNLSE in a Hamiltonian framework involving generalized definitions of momentum and energy operators. A central contribution of our study is the derivation of a fractional continuity equation and the identification of a broad class of solutions expressed in terms of Mittag-Leffler (ML) plane waves, which serve as the fractional generalization of standard exponential modes. A generalized Hamilton–Jacobi equation is derived from the classical velocity associated with ML plane waves, which bridges between fractional quantum dynamics and a classical action. Moreover, through a generalized Wick rotation, we introduced a connection between the FNLSE and a fractional Fokker–Planck equation describing a dual stochastic process. This formal mapping allows for a probabilistic reinterpretation of the dynamics and connects the behavior of FNLSE solutions to $Q$-Gaussian distributions ($Q=1-a$), which are characteristic of systems governed by non-extensive statistical mechanics. This stochastic correspondence opens potential avenues for future research in fractional quantum thermodynamics and transport processes in disordered media.

Our classification of solutions via separation of variables includes not only standard ML plane waves but also other chiral generalizations. Furthermore, the fractional Riccati method was used to construct a new family of bright and dark solitonic solutions, whose stability was investigated using a state distance metric that quantifies the robustness of these waveforms under perturbations. We found that, as the fractional parameter ($\alpha$) changes i.e. the memory effects are tuned, the bright solitons transform to dark solitons, which is equivalent to focusing-defocusing transition. The detailed table of Riccati-type solutions included in this paper serves as a reference for further analytical and numerical studies of soliton propagation in fractional media. In parallel, we applied the Adomian Decomposition Method to derive a series expansion of the FNLSE solutions, both in the case of chiral dynamics and for ML-based plane waves. This method proved to be particularly effective for approximating solutions in regimes where exact closed-form expressions are not readily available. Importantly, the Adomian framework naturally accommodates the nonlocal features of fractional derivatives, making it a valuable tool for semi-analytical treatment of similar models. Lastly, the Scaling Method allowed us to investigate self-similar solutions, leading to an understanding of how waveforms evolve under scale transformations. We derived explicit pure phase solutions originating from the ML phase structure, as well as hybrid solutions constructed through Adomian decomposition within the scaling ansatz. The existence of such self-similar profiles underscores the presence of underlying symmetries and conservation laws even in the fractional regime.\\

Taken together, the findings of this paper demonstrate the rich structure of the FNLSE and highlight the utility of fractional calculus in extending the toolkit of nonlinear wave theory. The combination of Riccati-based solitons, ML modes, stochastic mappings, and self-similar solutions reveals a coherent mathematical and physical picture of wave propagation in nonlocal, memory-bearing media. The experimental relevance of the FNLSE in physical systems such as optical lattices with long-range interactions, Bose-Einstein condensates with Lévy noise, or media with subdiffusive transport warrants investigation. Our analytical solutions could serve as testbeds for benchmarking experimental observations or guiding the design of systems where fractional nonlinear wave dynamics are dominant.

%==================================================

%%%%%%%%%%%%%%%%%%%%%%%%%%%%%%%%%%%%%%%%

%%%%%%%%%%%%%%%%%%%%%%%%%%%%%%%%%%%%%%
%%%%%%%%%%%%%%%%% Acknowledgments %%%%%%%%%%%%%%%%%%%%%

\section*{Authors' Contributions}
M.N.N. conceived the research idea and supervised the work. F. F. analyzed the data and performed the analytic calculations. F. F. and M.N.N. interpreted the results. F. F. performed the numerical calculations and prepared the initial draft of the manuscript, and M. N. N. contributed to writing and refining the final version of the manuscript.

%%%%%%%%%%%%%%%%%%%%%%%%%%%%%%%%%%%%%%%%%%%%%%%%%%%%%%%%%

\appendix

\section{Modified Riemann-Liouville Derivative}\label{App:RL}
In this section, some useful properties of modified RL are provided:
\begin{equation}
    D^\alpha_\zeta \zeta^\beta= \frac{\Gamma(\beta+1)}{\Gamma(\beta-\alpha+1)} \zeta^{\beta-\alpha},\quad \beta>0\label{eq.A1}
\end{equation}
\begin{equation}
   D^\alpha_\zeta c u(\zeta)=c D^\alpha_\zeta u(\zeta)\label{eq.A2}
\end{equation}

\begin{equation}
    D^{\alpha}_\zeta (u(\zeta)v(\zeta))=u(\zeta) D^{\alpha}_\zeta v(\zeta)+v(\zeta) D^{\alpha}_\zeta u(\zeta)\label{eq.A4}
\end{equation}
\begin{align}
    D^{\alpha}_\zeta u[v(\zeta)] &= u'_v(v) D^\alpha_\zeta v, \label{eq:A.5} \\
    &= D^\alpha_v u(v) (v'(\zeta))^\alpha. \label{eq:A6}
\end{align}

 Where c is a constant. The function ~\( u(\zeta) \) are non-differentiable in equation ~\ref{eq.A4} and ~\ref{eq:A.5} but becomes differentiable in \ref{eq:A6}. Similarly,~ \( v(\zeta) \) is non-differentiable in \ref{eq.A4}. Meanwhile, ~\( u(v) \) is differentiable in ~\ref{eq:A.5} but non-differentiable in ~\ref{eq:A6} \cite{jumarie2010approach}.The equations ~\ref{eq.A4}-~\ref{eq:A6} are derived directly from $D^\alpha_\zeta ~u(\zeta)\cong \Gamma(\alpha+1) D_\zeta u(\zeta)$.
Using these relations, the modified RL derivative of one-paramerer Mittag-Leffler function is calculated as
\begin{equation}
\begin{split}
   &\mathcal{D}^\alpha_x E_\alpha(i x^\alpha)=\sum_{k=0}^{\infty}\frac{\mathcal{D}^\alpha_x(i^k x^{\alpha k})}{\Gamma(\alpha k+1)}= \\&\sum_{k=1}^{\infty}\frac{i^k \Gamma(\alpha k+1) x^{\alpha(k-1)}}{\Gamma(\alpha k+1) \Gamma(\alpha(k-1)+1)}= \\&\sum_{s=0}^{\infty} \frac{i^{s+1} x^{\alpha s}}{\Gamma(\alpha s+1)}=i E_\alpha (i x^\alpha) \label{Mittag} 
\end{split}
\end{equation}
In a similar manner, using \eqref{eq:A6}, we have
\begin{equation}
\begin{aligned}
    \mathcal{D}^{\alpha}_\chi E_\alpha\left(i \left(\frac{w \chi^2}{2}\right)^\alpha\right)  
    &= \sum_{k=0}^{\infty} 
    \frac{i^k ~ \mathcal{D}^\alpha_\chi \left(\frac{w \chi^2}{2}\right)^{\alpha k}}
    {\Gamma(\alpha k +1)} \\
    &= w^\alpha \chi^\alpha \sum_{k=1}^{\infty} 
    \frac{i^k \left(\frac{w \chi^2}{2}\right)^{\alpha k-\alpha}}
    {\Gamma(\alpha k -\alpha +1)} \\
    &= i w^\alpha \chi^\alpha \sum_{s=0}^{\infty} 
    \frac{i^s \left(\frac{w\chi^2}{2}\right)^{\alpha s}}
    {\Gamma(\alpha s+1)} \\
    &= i w^\alpha \chi^\alpha 
    E_\alpha\left(i \left(\frac{w \chi^2}{2}\right)^\alpha\right).\label{AA7}
\end{aligned}
\end{equation}

In addition, Jumarie~\cite{jumarie2005representation} derived the following equality for the integral of any continuous function \( f \) with respect to \( d\zeta^\alpha \) 
\begin{equation}
\begin{split}
     \int_{0}^{x} f(\zeta) \, d\zeta^\alpha &= 
    \alpha \int_{0}^{x} (x-\zeta)^{\alpha-1} f(\zeta) \, d\zeta\\
    &=\Gamma(\alpha+1)\mathcal{J}_0 f(x),
    \quad 0<\alpha\leq 1.\label{A7}   
\end{split}
\end{equation}
Moreover, we have the following commutativity property:  
\begin{equation}
    \mathcal{D}^\alpha_t \mathcal{J}^\alpha_0 f(x,t) = \mathcal{J}^\alpha_0 \mathcal{D}^\alpha_t f(x,t).\label{A8}
\end{equation}
where the Riemann–Liouville integral and the Jumarie derivative are taken over the \( x \)- and \( t \)-domains, respectively. This holds because
\begin{equation}
\begin{split}
    \mathcal{D}^\alpha_t \mathcal{J}^\alpha_0 f(x,t) 
    &= \mathcal{D}^\alpha_t \left( \frac{1}{\Gamma(\alpha)} \int_{0}^{x} (x-\zeta)^{\alpha-1} f(\zeta,t) \, d\zeta \right) \\ 
    &= \frac{1}{\Gamma(\alpha)} \int_{0}^{x} (x-\zeta)^{\alpha-1} \mathcal{D}^\alpha_t f(\zeta,t) \, d\zeta \\ 
    &= \mathcal{J}^\alpha_0 \mathcal{D}^\alpha_t f(x,t).
\end{split}
\end{equation}

\section{Riccati Method}\label{SEC:Ricatti}
In this section, we present the various choices for \( F \), as follows:

\begin{itemize}
    \item \textbf{C1:} $A = 1$ and $B = 1$: $F = \tan(\xi, \alpha)$.
    \item \textbf{C2:} $A = -1$ and $B = -1$: $F = \cot(\xi, \alpha).$ 
    \item \textbf{C3:} $A = 1$ and $B = -1$: $ F = \tanh(\xi, \alpha)$, $F = \coth(\xi, \alpha)$.
    \item \textbf{C4:} $A = \dfrac{1}{2}$ and $B = -\dfrac{1}{2}$: $F = \coth(\xi, \alpha)+\text{csch}(\xi, \alpha)$.
    \item  \textbf{C5:} $A = \dfrac{1}{2}$ and $B = \dfrac{1}{2}$: $F = \csc(\xi, \alpha)-\tan(\xi, \alpha)$.
    \item  \textbf{C6:} $A = 1$ and $B = -4$, $F = \dfrac{\tanh(\xi, \alpha)}{1+\tanh^2(\xi, \alpha)}$.
    \item \textbf{C7:} $A = 1$ and $B = 4$: $F = \dfrac{\tan(\xi, \alpha)}{1-\tan^2(\xi, \alpha)}$.
    \item  \textbf{C8:} $A = -1$ and $B = -4$: $F = \dfrac{\cot(\xi, \alpha)}{1-\cot^2(\xi, \alpha)}$.
\end{itemize}
 where the generalized hyperbolic and trigonometric functions are defined in table \ref{table2}.
\begin{table}[H]
\centering
\renewcommand{\arraystretch}{1.8}
\begin{tabular}{|c|c|}
\hline
\textbf{Function} & \textbf{Definition} \\
\hline
\(\cosh(\xi, \alpha)\) & \(\displaystyle \frac{E_{\alpha}(\xi^{\alpha}) + E_{\alpha}(-\xi^{\alpha})}{2} \) \\[4pt]
\(\sinh(\xi, \alpha)\) & \(\displaystyle \frac{E_{\alpha}(\xi^{\alpha}) - E_{\alpha}(-\xi^{\alpha})}{2} \) \\[4pt]
\(\cos(\xi, \alpha)\)  & \(\displaystyle \frac{E_{\alpha}(i \xi^{\alpha}) + E_{\alpha}(-i \xi^{\alpha})}{2} \) \\[4pt]
\(\sin(\xi, \alpha)\)  & \(\displaystyle \frac{E_{\alpha}(i \xi^{\alpha}) - E_{\alpha}(-i \xi^{\alpha})}{2i} \) \\[4pt]
\(\tanh(\xi, \alpha)\) & \(\displaystyle \frac{\sinh(\xi, \alpha)}{\cosh(\xi, \alpha)} \) \\[4pt]
\(\tan(\xi, \alpha)\)  & \(\displaystyle \frac{\sin(\xi, \alpha)}{\cos(\xi, \alpha)} \) \\[4pt]
\(\coth(\xi, \alpha)\) & \(\displaystyle \frac{1}{\tanh(\xi, \alpha)} \) \\[4pt]
\(\cot(\xi, \alpha)\)  & \(\displaystyle \frac{1}{\tan(\xi, \alpha)} \) \\[4pt]
\(\text{sech}(\xi, \alpha)\) & \(\displaystyle \frac{1}{\cosh(\xi, \alpha)} \) \\[4pt]
\(\sec(\xi, \alpha)\) & \(\displaystyle \frac{1}{\cos(\xi, \alpha)} \) \\[4pt]
\(\text{csch}(\xi, \alpha)\) & \(\displaystyle \frac{1}{\sinh(\xi, \alpha)} \) \\[4pt]
\(\csc(\xi, \alpha)\) & \(\displaystyle \frac{1}{\sin(\xi, \alpha)} \) \\
\hline
\end{tabular}
\caption{Generalized trigonometric and hyperbolic functions using the Mittag-Leffler function}\label{table2}
\end{table}

\section{Adomian Decomposition Method}\label{SEC:AdomianApp}
\appendix
\section*{Appendix C: Adomian Polynomials}

In the Adomian Decomposition Method (ADM), the solution \( u \) of a nonlinear problem is expressed as an infinite series:
\begin{equation}
    u = \sum_{n=0}^\infty u_n,
\end{equation}
where each \( u_n \) is a component function to be determined. Correspondingly, the nonlinear term \( \mathcal{N}(u) \) is decomposed into a series of Adomian polynomials:
\begin{equation}
    \mathcal{N}(u) = \sum_{n=0}^\infty A_n,
\end{equation}
where the terms \( A_n \) are the Adomian polynomials, defined by:
\begin{equation}
A_n = \frac{1}{n!} \left. \frac{\text{d}^n}{\text{d}\lambda^n} \mathcal{N}\left(\sum_{j=0}^n u_j \lambda^j\right) \right|_{\lambda=0}. \label{App:C1}
\end{equation}

This formula allows for the systematic calculation of the \( A_n \) terms, where \( \lambda \) is an auxiliary parameter introduced for the expansion, and \( \mathcal{N} \) is assumed to be a nonlinear operator. In our case, it is given by \[
\mathcal{N}(u) = |u|^a u,
\] where \( a \) is a real parameter. Below, we provide several components of the corresponding Adomian polynomials.

\begin{equation}
    \begin{aligned}
        A_0 &= |u_0|^a u_0, \\
        A_1 &= \left( \frac{a}{2} + 1 \right) |u_0|^a u_1 + \frac{a}{2} u_0^{\frac{a}{2} + 1} \overline{u_1} \overline{u_0}^{\frac{a}{2} - 1}, \\
        A_2 &= \frac{a}{4} \left( \frac{a}{2} + 1 \right) u_0^{\frac{a}{2} - 1} u_1^2 \overline{u_0}^{\frac{a}{2}} + \left( \frac{a}{2} + 1 \right) \mid{u_0}\mid^{a} u_2  \\
        &\quad + \frac{a}{2} \left( \frac{a}{2} + 1 \right) u_0^{\frac{a}{2}} u_1 \overline{u_1} \overline{u_0}^{\frac{a}{2} - 1} + \frac{a}{2} u_0^{\frac{a}{2} + 1} \overline{u_2} \overline{u_0}^{\frac{a}{2} - 1} \\
        &\quad + \frac{a}{4} \left( \frac{a}{2} - 1 \right) u_0^{\frac{a}{2} + 1} \overline{u_1}^2 \overline{u_0}^{\frac{a}{2} - 2},\\
        &\dots
    \end{aligned}
\end{equation}
Substituting these decompositions into the original equation yields a recurrence relation, which can be used to compute the components \( u_n \) sequentially.
\section{Mapping FNLSE to a fractional non-linear Fokker-Planck Equation} \label{AppD}

As we mentioned in SEC.~\ref{SEC:Wicksec}, using a generalized Wick's rotation one finds
\begin{equation}
\mathcal{D}_\tau^{\alpha}P(x,\tau)=q\mathcal{D}_x^{2\alpha}P(x,\tau)+qP(x,\tau)^{a+1}.\label{D1}
\end{equation}
We then try the trial function $P(x,\tau)=P_0(x,\tau)Q(x,\tau)$, where $P_0$ is an auxiliary function that satisfies 
\begin{equation}
\mathcal{D}_\tau^\alpha P_0(x,\tau)=qQ(x,\tau)^aP_0(x,\tau)^{a+1}.
\label{Eq:P0}
\end{equation}
Then the Eq.~\ref{D1} casts to
\begin{equation}
\mathcal{D}_\tau^{\alpha}Q(x,\tau)=\frac{q}{P_0(x,\tau)}\mathcal{D}_x^{2\alpha}\left(P_0(x,\tau)Q(x,\tau)\right), \label{D3}
\end{equation}
which is a generalized Fokker-Planck equation for $Q$. For $\alpha=1$, the solution of Eq.~\ref{Eq:P0} is ($a\ne 0$):
\begin{equation}
P_0(x,\tau)^{-a}=P_0(x,\tau_0)^{-a}-qa\int_{\tau_0}^{\tau}Q(x,\tau')^a\text{d}\tau',
\label{Eq:P0int}
\end{equation}
while for $a=0$ the solution is
\begin{equation}
P_0(x,\tau)|_{a\to 0}=P_0(x,\tau_0)\exp [q(\tau-\tau_0)],
\end{equation}
where $\tau_0$ is a reference point. One may try the scaling relation 
\begin{equation}
Q(x,\tau)\equiv\tau^{-\gamma}F\left(\frac{x}{\tau^{\gamma}}\right) \label{D6}
\end{equation}
to evaluate the integral Eq.~\ref{Eq:P0int}, where $F(y)$ is a slow-varying function, and $\gamma$ is an exponent related to $\alpha$ and $a$. Defining $\zeta\equiv \frac{x}{\tau^\gamma}$, so that $\text{d}\tau=-\frac{x^\gamma}{\gamma}\zeta^{-(\gamma+1)}\text{d}\zeta$, one finds
\begin{equation}
\int_{\tau_0}^{\tau}F(x,\tau')^a\text{d}\tau'\approx \frac{\bar{F}^a}{\gamma(a-\gamma)}\left(\frac{1}{\tau_0^{\gamma(a-\gamma)}}-\frac{1}{\tau^{\gamma(a-\gamma)}}\right),
\end{equation}
where $\bar{F}$ is the average of $F$ in the integration interval. Incorporating this into Eq.~\ref{Eq:P0int} gives
\begin{equation}
P_0(x,\tau)=\frac{P_0(x,\tau_0)}{\left(1-Ca(\tau_0^{-\eta}-\tau^{-\eta})\right)^{\frac{1}{a}}},
\label{Eq:QGaussian}
\end{equation}
where
\begin{equation}
C\equiv \frac{q\bar{F}^a}{\gamma(a-\gamma)P_0(x,\tau_0)^a}, \ \eta\equiv \gamma(a-\gamma).
\end{equation}
Equation~\ref{Eq:QGaussian} is a $Q$-Gaussian function:
\begin{equation}
P_0(x,\tau)=\frac{P_0(x,\tau_0)}{\mathbb{E}_{Q}\left(-C(\tau_0^{-\eta}-\tau^{-\eta})\right)},
\label{Eq:QGaussian}
\end{equation}
where $Q=1-a$ and
\begin{equation}
\mathbb{E}_{Q}(y)\equiv \left(1+(1-Q)y\right)^{\frac{1}{1-Q}}.
\label{Eq:QGaussian}
\end{equation}
Given the function $P_0(x,\tau)$, one finds the following Fokker-Planck equation:
\begin{equation}
    \mathcal{D}_\tau^\alpha f(x,\tau)=\frac{p}{P_0(x,\tau)}\mathcal{D}_x^{2\alpha}\left(P_0(x,\tau)f(x,\tau)\right).
\end{equation}
\section{proof of a solution to nonlinear Equation}\label{AppE}
In this section, we demonstrate that the solution of the following fractional non-linear is actually a $Q$-Gaussian function. The equation of interest is:
\begin{equation}
    \mathcal{D}^\zeta_x f = \beta f^{\gamma}.
\end{equation}
Taking the integral with respect to \( dx^\alpha \) yields
\begin{equation}
    \int \frac{\mathcal{D}^\zeta_x f}{f^\gamma} \, \text{d}x^\zeta = \beta \int \text{d}x^\zeta, \label{E2}
\end{equation}
where $\text{d}^\zeta x$ has already been defined in terms of the fractional integrals Eq.~\ref{Eq:RLIntegral}.
 We then use the identity
\begin{equation}
    \mathcal{D}^\zeta_x f^{1-\gamma} = (1-\gamma) f^{-\gamma} \mathcal{D}^\zeta_x f,
\end{equation}
which gives rise to
\begin{equation}
    \frac{\mathcal{D}^\zeta_x f}{f^\gamma} = \frac{1}{1-\gamma} \mathcal{D}^\zeta_x f^{1-\gamma}. \label{E4}
\end{equation}
Substituting equation \eqref{E4} into equation \eqref{E2}, we obtain
\begin{equation}
    \frac{1}{1-\gamma} \left( f^{1-\gamma} - f_0^{1-\gamma} \right) = \beta  \left( x^\zeta - x_0^\zeta \right),
\end{equation}
where $x_0$ and $f_0$ are some initial quantities. Solving this for \( f \), we get
\begin{equation}
    f = f_0 \mathbb{E}_\gamma\left[\frac{\beta}{f_0^{1-\gamma}}\left(x^\zeta - x_0^\zeta\right)\right], \label{functionf}
\end{equation}
 where $\mathbb{E}_Q$ is defined in~\ref{Eq:QGaussian}.

 \section{Stability of a Soliton} \label{Appendix:stability}
Suppose that for the FNLSE Eq.~\ref{Eq:NLSE} we have a solution with the parameter $\theta$, i.e.
 \begin{equation}
u=u_\theta(x,t).
 \end{equation}
where $\theta$ is the free parameter (or a set of free parameters) that classifies the solutions. Specifically, for $\theta=\theta^*$ the solution is soliton.
A stability test is comprised of a slight change of the solution:
\begin{equation}
u\equiv u_{\theta^*+\delta\theta}\approx u_{\theta^*}+\delta u,
\end{equation}
where $\delta\theta$ is very small perturbation, and track how the modified solution changes upon time. More precisely, one has to check if the solution of the following equation decays to the original soliton one:
\begin{equation}  
    i \mathcal{D}_t^{\alpha} u_{\theta^*+\delta\theta} - p \mathcal{D}_x^{\beta} u_{\theta^*+\delta\theta} - q |u_{\theta^*+\delta\theta}|^a u_{\theta^*+\delta\theta} = 0. 
    \label{Eq:NLSE2}  
\end{equation}
To the first order of $\delta u$, and using the Eq.~\ref{Eq:NLSE} for $u_{\theta^*}$, one can linearize the equation governing $\delta u$:
\begin{equation}
i \mathcal{D}_t^{\alpha} \delta u - p \mathcal{D}_x^{\beta} \delta u - qV_a(x,t) \delta u = 0,
\end{equation}
where $\bar{u}$ is complex conjugate of $u$, and 
\begin{equation}
V_a(x,t)\equiv    \left[1+\frac{a}{2}\left(\frac{u_{\theta^*}(x,t)}{\bar{u}_{\theta^*}(x,t)}+1\right)\right] \left|u_{\theta^*}(x,t)\right|^a
\end{equation}
is a complex time-dependent effective potential. One then has to find the solution of the above equation and observe if $\delta u$ decays over time. Note that for the linear Schrodinger equation ($a=0$), $V_a(x,t)=1$. \\

An equivalent method to address stability is to directly test whether the perturbed solution comes back to the soliton one over time or not. To this end, we define the distance
\begin{equation}
d(t)\equiv\int_{-\infty}^\infty\text{d}x\left|u_{\theta^*+\delta \theta}(x,t)-u_{\theta^*}(x,t)\right|.
\label{Eq:stability}
\end{equation}
For a stable solution, $d(t)$ should decay with time.

\begin{figure*}[t]
\centering

\begin{subfigure}[b]{0.4\linewidth}
    \centering
    \includegraphics[width=\linewidth]{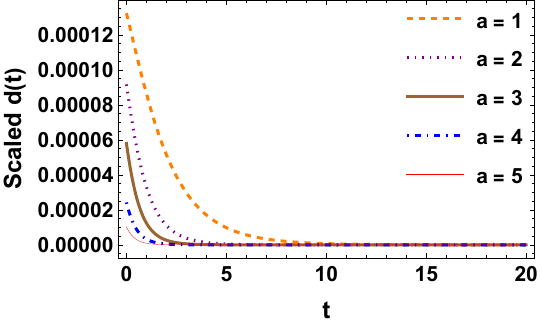}
    
    (a)
\end{subfigure}
\begin{subfigure}[b]{0.4\linewidth}
    \centering
    \includegraphics[width=\linewidth]{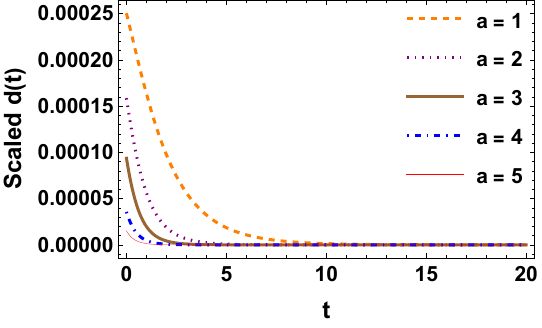}
    
    (b)
\end{subfigure}

\vspace{0.5cm}

\begin{subfigure}[b]{0.4\linewidth}
    \centering
    \includegraphics[width=\linewidth]{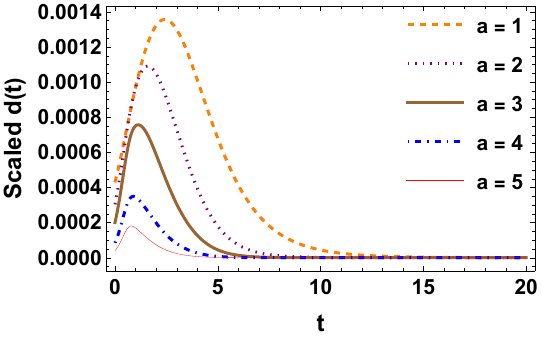}
    
    (c)
\end{subfigure}
\begin{subfigure}[b]{0.4\linewidth}
    \centering
    \includegraphics[width=\linewidth]{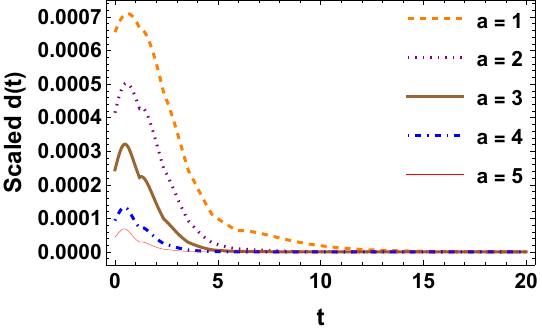}
    
    (d)
\end{subfigure}

\caption{Stability test  of $u_7$ with $r=0.5$, and $\alpha = 0.8$. 
(a) $p=1$, $q=1$; 
(b) $p=1$, $q=-1$; 
(c) $p=-1$, $q=1$; 
(d) $p=-1$, $q=-1$.}
\label{fig:stability4}
\end{figure*}

\begin{figure*}[t]
\begin{subfigure}{0.4\linewidth}
    \includegraphics[width=\linewidth]{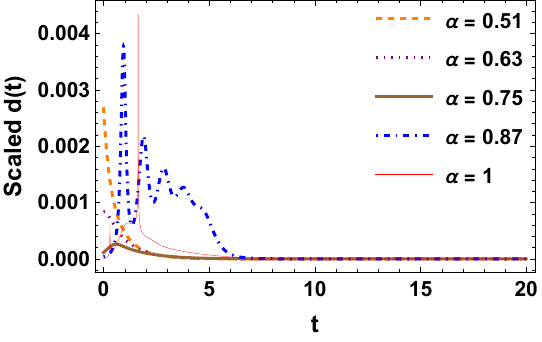}
    (a)
\end{subfigure}
\begin{subfigure}{0.4\linewidth}
    \includegraphics[width=\linewidth]{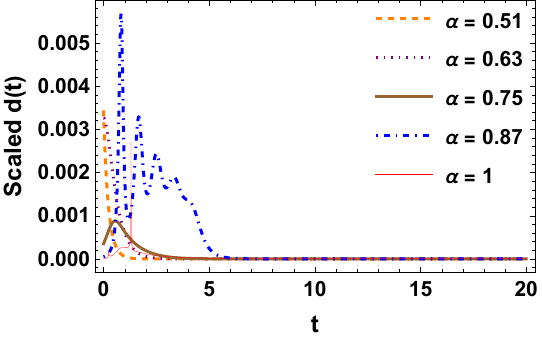}
    (b)
\end{subfigure}
\caption{Stability test with $r=0.5$, $p=-1$, $q=-1$, and $\alpha = 0.8$. 
(a) $u_4$; 
(b) $u_5$;} 

\label{fig:stability5}
\end{figure*}

\begin{figure*}[t]
\centering

\begin{subfigure}[b]{0.4\linewidth}
    \centering
    \includegraphics[width=\linewidth]{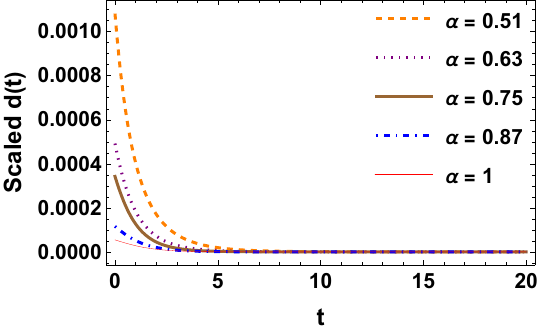}
    
    (a)
\end{subfigure}
\begin{subfigure}[b]{0.4\linewidth}
    \centering
    \includegraphics[width=\linewidth]{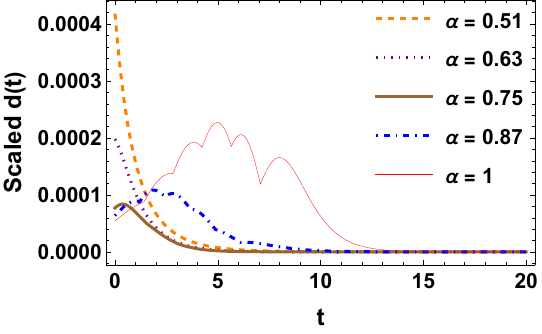}
    
    (b)
\end{subfigure}

\vspace{0.5cm}

\begin{subfigure}[b]{0.4\linewidth}
    \centering
    \includegraphics[width=\linewidth]{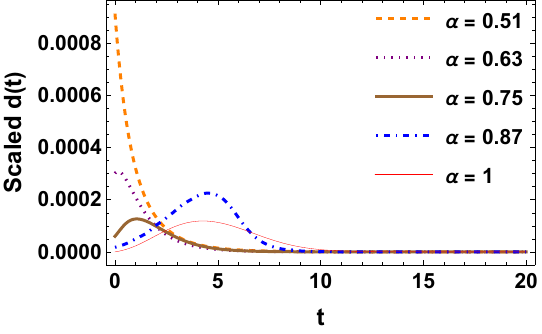}
    
    (c)
\end{subfigure}
\begin{subfigure}[b]{0.4\linewidth}
    \centering
    \includegraphics[width=\linewidth]{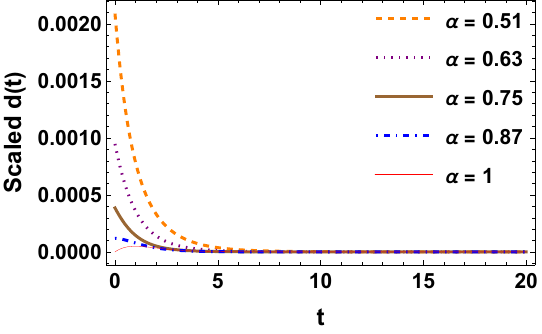}
    
    (d)
\end{subfigure}

\caption{Stability test  of $u_7$ with $r=0.5$, and $\alpha = 0.51,0.63,0.75,0.87,1$. 
(a) $p=1$, $q=1$; 
(b) $p=-1$, $q=-1$; 
(c) $p=-1$, $q=1$; 
(d) $p=1$, $q=-1$.}
\label{fig:stability6}
\end{figure*}

Figures illustrating the stability behavior for a range of \( p \), \( q \), and \( a \) values are provided below. As shown in Figs.~\ref{fig:stability4}, \ref{fig:stability5}, and \ref{fig:stability6}, the quantity \( d(t) \) decays over time, indicating that the solutions become increasingly stable as time progresses.

%===================
\begin{figure*}[t]
\begin{subfigure}{0.4\linewidth}
    \includegraphics[width=\linewidth]{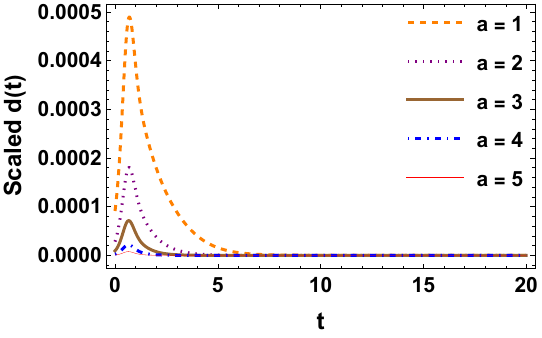}
    (a)
\end{subfigure}
\begin{subfigure}{0.4\linewidth}
    \includegraphics[width=\linewidth]{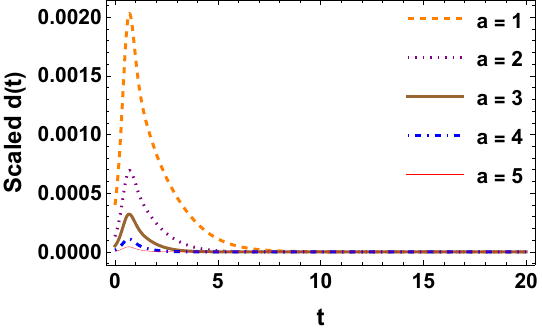}
    (b)
\end{subfigure}
\caption{Stability test with $r=0.5$, $p=-1$, $q=-1$, and $\alpha = 0.8$. 
(a) $u_4$; 
(b) $u_5$;} 

\label{fig:stability3}
\end{figure*}
%================
%=================
%==================
%===================
\bibliography{refs}

\end{document}